\documentclass[reqno,10pt]{amsart}
\usepackage{amsmath,amsfonts,amsthm,amssymb}
\usepackage{times}
\numberwithin{equation}{section}

\def\N{\mathbb N}
\def\R{\mathbb R}

\def\E{\mathbb E}  

\def\dist{\operatorname{dist}}   
\def\dim{\operatorname{dim}}  

\def\too#1{\mathop {\longrightarrow}_{#1}}  



\renewcommand{\epsilon}{\varepsilon}
\newcommand{\x}{\boldsymbol{x}}
\newcommand{\X}{\boldsymbol{X}}
\renewcommand{\k}{\boldsymbol{k}}
\newcommand{\ip}[2]{\langle #1 , #2 \rangle} 

\def\qed{$\Box$}
\def\sig{ \sigma}
\newcommand{\eq}[1]{eq.~(\ref{#1})}
\newcommand{\Ev}[1]{\E \left( #1 \right) }

\newtheorem{thm}{Theorem}[section]

\newtheorem{Corollary}[thm]{Corollary}
\newtheorem{cor}[thm]{Corollary}
\newtheorem{lem}[thm]{Lemma}

\newtheorem{defn}[thm]{Definition}

\renewcommand{\epsilon}{\varepsilon}

\begin{document}
\flushbottom

\title{Mean-Field Spin Glass models from the Cavity--ROSt Perspective}

\author{Michael Aizenman} %
              \address{M. Aizenman, Departments of Mathematics and Physics,
             Princeton University, Princeton NJ 08544}
              \email{aizenman@princeton.edu}%

 \author{Robert Sims}
\address{R. Sims, Department of Mathematics, University of California at Davis,
Davis CA 95616} \email{rjsims@math.ucdavis.edu}%

\author{Shannon L. Starr} %
\address{S. L. Starr, Department of Mathematics,
             UCLA, Box 951555, Los Angeles, CA 90095}%
\email{sstarr@math.ucla.edu}%

\date{July 25, 2006 \\    
\indent Based on talks given at `Young Res. Symp.', Lisbon 2003, and  `Math. Phys. of Spin Glasses', Cortona 2005.}
\maketitle

\begin{abstract}
The Sherrington-Kirkpatrick spin glass model has been studied 
as a source of insight into the statistical mechanics of systems 
with highly diversified collections of competing low energy states. 
The goal of this summary is to present some of the ideas which 
have emerged in the mathematical study of its free energy.  In
particular, we highlight the perspective of the cavity dynamics,
and the related variational principle.  These are expressed in
terms of Random Overlap Structures (ROSt), which are used to
describe the possible states of the reservoir in the cavity step.
The Parisi solution is presented as reflecting the ansatz that it
suffices to restrict the variation to hierarchal structures which
are discussed here in some detail. While the Parisi solution was
proven to be correct, through recent works of F. Guerra and M.
Talagrand, the reasons for the effectiveness of the Parisi ansatz
still remain to be elucidated.  We question whether this could be
related to the quasi-stationarity of the special subclass of
ROSts given by Ruelle's hierarchal  `random probability cascades' 
(also known as GREM).
\end{abstract}


\setcounter{tocdepth}{1}
 \tableofcontents
\newpage
\addtocounter{section}{-1}

\section{An outline}

The Sherrington-Kirkpatrick spin glass model has been studied as a
source of insight into statistical mechanics of systems with
highly diversified collections of patterns for the minimization of
the free energy, or energy.  The model is based on a  Hamiltonian
which incorporates interactions with high levels of frustration
and disorder. The goal of this article is to present some of the
ideas which have emerged in the study of the SK model, and in
particular highlight an approach for the analysis of its free
energy influenced by the cavity perspective.

The discussion is organized as follows.

In Section 1 we present the Sherrington-Kirkpatrick
model~\cite{SK}, and comment on some of its basic features and
puzzles.   A more general version of the model is  presented in
Appendix C.  Among the essential features exhibited by these
models is the presence of  rich diversity of low energy
configurations.   A proposal for a solution of the SK model was
developed in a series of works, driven by the astounding insight
of G. Parisi~\cite{P}. An essential feature of the proposed
solution is the ansatz that at low temperatures the model's Gibbs
states exhibit a hierarchal structure.  The Parisi approach was
further clarified by Mezard, Parisi and Virasoro~\cite{MPV}, and
proceeding through Derrida's REM and GREM calculations~\cite{Der}
which have in turn motivated Ruelle's construction~\cite{Ru}.

In Section 2 we present the cavity perspective and show that it
naturally leads to the random overlap structure as the order
parameter. An order parameter is a quantity which captures an
essential feature of the system, whose determination provides key
information on the system's state.  In the ferromagnetic Ising
model the role is usually played by the magnetization. Parisi has
presented his solution of the SK spin glass  model as involving an
order parameter which is a monotone function of the unit interval.
That, however, presupposes `ultrametricity' or the hierarchal
structure discussed below.  Without such an assumption, we argue
that the natural order parameter is a ROSt.

Section 3 presents interpolation techniques.  The considerable
recent progress in the mathematical study of the SK model was
stimulated, and indeed enabled, by the interpolation argument
which was introduced in the work of F. Guerra and F. L. 
Toninelli~\cite{GT}.  The basic tools are presented here within
the context of ROSt.

In Section 4 we discuss a variational formulation of the
solution~\cite{AS2}, which starts with an extension to general
ROSts of the remarkable statement  of F. Guerra~\cite{Gue} that
Parisi's ansatz provides a rigorous bound.  The extended
variational principle is shown to provide the correct answer, but is not computationally effective.

In part 5 we present a hierarchal ``random probability cascade''
(RPC) model, following closely a construction which was formulated by D. Ruelle.   For ROSt within this class, the variational quantity can be presented as a functional defined over monotone functions (equivalently, probability measures) over the unit interval.  The Parisi solution can be explained as based on the ansatz that it suffices to restrict the variation to ROSts in that class.

The Parisi expression for the free energy was proved by M.
Talagrand to be correct~\cite{Tal_AM}.  This was established through
the criterion which is provided by Guerra's interpolation bound.
However,  the notable  result still does not fully address the
challenge of explaining the reasons for the validity of the
Parisi  ansatz.   In part 6, we  comment on that, and on the
question whether the reasons for the validity of Parisi's ansatz
could be related to the remarkable quasi-stationarity of the
hierarchal RPC  under the dynamical process which is naturally
associated with the cavity picture~\cite{RA}.


\section{The  Sherrington-Kirkpatrick spin glass model: Basics}

\subsection{Formulation of the model}

Spin glass models were formulated in an attempt to provide
analyzable and instructive examples of systems with intricate
dynamics and equilibrium states of rich structure. A prime example
is the Sherrington-Kirkpatrick model \cite{SK}, whose
configurations are described by $N$ spin variables $\{
\sigma_i\}_{i=1,...,N}$ taking values  $\pm 1$, and interacting
via the random Hamiltonian:
\begin{equation} \label{H}
-H_N (\sigma;\omega, h) \ = \  \frac{1}{\sqrt{N}} \sum_{i<j}
J_{ij}(\omega) \sigma_i \sigma_j  + h \sum_i \sigma_i \, ,
\end{equation}
where $J_{ij}(\omega)$ are iid  gaussian random variables
with normal distribution, and $h$ is a real parameter.

Since its introduction, the model has attracted considerable
discussion and shown itself to contain various surprises.
Even before one addresses the complex structure which the
model aims to express, one encounters certain basic questions
for which the answer is  not immediate.

\subsection{Comments on the ground state energy}

The normalizing factor $N^{-1/2}$, included in \eq{H}, ensures
that the lowest energy
$$\mathcal{E}_N(\omega,h) = \min_{\sigma} H_N (\sigma;\omega, h)  $$
is typically of order $N$, even when $h=0$. However, the fact that
this is so requires an argument, since for any a-priori chosen
configuration the typical order of magnitude of the energy is only
$O(N^{1/2})$ (the distribution of the collection of energies is
symmetric under reflection, due to the invariance of the
distribution of $\{ J\}$ under:  $J_\cdot \to -J_\cdot$.)
The resolution of this issue is easy.
However, some of the next questions are not so simple: \\

  [Q 1.] Is the random variable $\mathcal{E}_N(\omega,h) / N$
sharply distributed, for high $N$, and does it converge in
distribution to a constant as $N\to \infty$?  \\

The answer to both questions is "yes" (see~\cite{Tal} and
references therein) though for the second question it took
considerable time for the answer to be established rigorously.
The task was accomplished in a clever and simple argument of
Guerra and Toninelli~\cite{GT}.   \\

  [Q 2.]  Compute, or give an effective way to estimate,  the
  distributional limit (which does exist),
\begin{equation} \nonumber E_o \, \stackrel{ \mathcal{D}}{=} \,
\lim_{N\to \infty} \mathcal{E}_N(\omega,h) /N \quad .
\end{equation} \\

 [Q 3.] Produce an algorithm for determining the energy minimizing
 configuration, or at least find one for which the resulting energy
per volume $H_N(\sigma; \omega)/N $ is close to $E_o$. \\

It turns out that in order to determine $E_o$, either theoretically
or numerically, it is essential to consider the equilibrium states
of the model at positive temperatures, which  are, of course,
of intrinsic interest.  Thus,
one is led to consider: \\
$ \bullet$  the {\em partition function},
$$
Z_N(\beta; \omega, h) \ = \  \sum_{\sigma}
e^{-\beta H_N (\sigma;\omega, h)} \, ,
$$
$ \bullet$   the {\em quenched free energy}, which is $(-\beta)$
times the following  quantity
$$
Q_N(\beta,h) \ = \ \E(\log Z_N(\beta;\omega, h) ) \,
$$
where $\E(\cdot)$ represents the average over the random `environment' $\omega$.

A derivative of the free energy yields the mean value of the
energy density in the quenched state:
\begin{equation}
E_N(\beta,h) \ := \ \frac{1}{N} \E( \mathcal{E}_N(\beta;\omega, h)) \
= \  - \frac{\partial}{\partial \beta} Q_N(\beta,h) / N \, ,
\end{equation}
where $\mathcal{E}_N(\beta;\omega, h) $  is the Gibbs state average energy density
\begin{equation} \label{gibbs}
\mathcal{E}_N(\beta;\omega, h) \ = \    \sum_{\sigma} H_N (\sigma;\omega, h) \,
\frac{e^{-\beta H_N (\sigma;\omega, h)}}{Z_N(\beta; \omega, h)}    \, .
\end{equation} Standard convexity arguments imply that the mean is, at
almost every inverse- temperature $\beta$, also the typical value
of the energy density. More can be said on the so called {\em self averaging}
property of $\mathcal{E}_N(\beta;\omega, h)$ through the `concentration of measure'
principle which is nicely presented in the book of Talagrand~\cite{Tal}.

\subsection{Diversity}

Before we turn to the more detailed discussion of the free energy,
let us comment on the question concerning explicit algorithms for
finding low energy configurations. Two natural algorithms, which
are discussed in greater
detail in \cite{ALR} for $h=0$, are: \\
{\em i.} The  {\em greedy algorithm}:  $\sigma_i$
is determined successively, with respect to some order of the indices,
 by optimizing at each step the
sign of the contribution of the new terms.  For instance,
\begin{equation}
\sigma_1=+1\, ,  \qquad \mbox{ and for $i=2, ...., N$:} \quad
 \sigma_i = - \mbox{sign}\left\{ \sum_{j=1,...,i-1} J_{ij} \sigma_j \right\} \, .
\end{equation}
{\em ii.} The {\em eigenstate-shadowing} algorithm:
\begin{equation}
\sigma_i = \mbox{sign} \psi_i(\omega)
\end{equation}
with $\psi_i(\omega) $  one of the lowest eigenstates of the Hermitian matrix
$J_{ij}(\omega) $, which is sampled with the GOE distribution.  \\
A point to be appreciated here is that while the typical spectrum of the
corresponding quadratic form is known, through Wigner's celebrated
semi-circle law, the non-linear problem of determining the minimum
restricted to the vertices of the hypercube in $\R^N$ is, at present,
much harder.

The greedy algorithm, which typically yields configurations with
$H_N(\sigma; \omega, 0)/N \approx - 0.5319$ can be easily improved upon,
while an improvement over the second one, which typically yields
$H_N(\sigma; \omega, 0)/N \approx - 0.6366$ \cite{ALR},  presents a harder challenge.

It may be noted that both algorithms allow for the construction
of many very different configurations with comparable energy.
This does not yet prove that such a diversity persists at the bottom
of the spectrum, since neither yields the ground state energy per spin,
but nevertheless the diversity seen here does offer
a hint of the diversity characteristic of the model.
Those observations lead one to the fascinating question: \\

[Q 4.] How much variety is there among the low energy configurations? \\

By flipping a few spins of the minimizing
configuration, one can produce many configurations with
energy in the range $\mathcal{E}_N(\omega,h)+O(1)$.
However, the question is whether one finds configurations with energies
close to the ground state which are extensively distinct from each other.
For this purpose,  the distance between the
configurations  may be expressed through their  overlap, which is defined as
\begin{equation}
q_{\sigma,\sigma'} \ = \ \frac{1}{N} \sum_{i=1}^N  \sigma_i  \sigma_i'  \, ,
\end{equation}
with $\dist(\sigma,\sigma') := 1-q_{\sigma,\sigma'}^2$.  According to
the Parisi picture,  at the bottom of the
spectrum one should find a diverse collection  of ``competing''
configurations, whose energies and overlaps resemble  a RPC process.
The description is slightly complicated by the need to lump the configurations
into equivalence classes, according to their mutual overlaps.  Furthermore,
the discussion of the ground state is (so far) accessible only after understanding
the structure of the positive temperature Gibbs equilibrium states.


\section{The Cavity Perspective} \label{cav_perspect}

\subsection{The incremental free energy} \mbox{ }

It is convenient to present the pressure $P_N(\beta,h)\, :=\,
Q_N(\beta,h)/N$  as a sum of increments, which describe the effect
of the gradual increase in the  system's size, starting from $Z_0
=\equiv 1$:
\begin{equation}\label{eq:telescop}
P_N(\beta,h)\, :=\, \frac{1}{N} \E(\ln\, Z_N)\, =\, \frac{1}{N} \sum_{n=0}^{N-1} \E\Big(\ln\, \frac{Z_{n+1}}{Z_n}\Big)\, . \end{equation}
The sequence $P_N(\beta,h)$ converges  if and only if the sequence of increments is Cesaro-convergent, in which case:
\begin{equation}\label{eq:Cesaro_limit}
        P(\beta,h)\, :=\, \lim_{N\to\infty} P_N(\beta,h)\, =\, \textrm{(c--)} \lim_{N \to \infty} \E\Big(\ln\, \frac{Z_{N+1}}{Z_N}\Big)\, .
\end{equation}

For an intuitive description of the incremental term let us
describe the configuration of a large reservoir of $N$ spins by
the symbol $\alpha = (\sigma_1,\dots,\sigma_N)$, and let the next
spin be denoted $\hat{\sigma} = \sigma_{N+1}$. Then
\begin{equation}
        \frac{Z_{N+1}}{Z_N}\, =\, \frac{\sum_{\alpha,\hat{\sigma}} e^{-\beta H_{N+1}(\alpha,\hat{\sigma};\omega)}}{\sum_{\alpha} e^{-\beta H_N(\alpha;\omega)}}\, .
\end{equation}
We would like to cast the ratio as the effect of the addition of the single spin $\hat{\sigma}$ to a reservoir whose state is described by $\alpha$, and is governed by the Hamiltonian
$H_N(\alpha, \omega)$ .
First, however, one needs to deal with a minor inconvenience: as we go from size $N$ to $N+1$, the interaction in $\alpha$ diminishes because of the change in the normalizing factor
\begin{equation}
        \frac{1}{\sqrt{N}} J_{ij}\, \longrightarrow\, \frac{1}{\sqrt{N+1}} J_{ij}\, .
\end{equation}
To address this, we rewrite the interaction $H_N(\alpha)$ in a form which will allow a natural subtraction:
\begin{equation} \label{eq:new_N}
-H_N(\alpha)\, \stackrel{\mathcal{D}}{=}\, \frac{1}{\sqrt{N+1}} \sum_{i<j}^{N} J_{ij} \alpha_i \alpha_j
+ \underline{h} \cdot \underline{\alpha} + \frac{1}{\sqrt{N(N+1)}} \sum_{i<j}^{N} \tilde{J}_{ij} \alpha_i \alpha_j\, ,
\end{equation}
with  $\tilde{J}_{ij}$ are independent normal Gaussians, and
$\underline{h}$ is the vector with all components equal $h$. This
is to be compared with
\begin{equation} \label{eq:new_Nplus}
-H_{N+1}(\alpha,\hat{\sigma})\, =\,\frac{1}{\sqrt{N+1}} \sum_{i<j}^{N} J_{ij} \alpha_i \alpha_j
+ \underline{h} \cdot \underline{\alpha} + \frac{1}{\sqrt{N+1}}
\sum_{i=1}^N J_{i,N+1} \alpha_i \hat{\sigma} + h \hat{\sigma}\, ,
\end{equation}

For brevity, let us denote the two terms which appear above as
independent additions to $H_N(\alpha)$, as the Gaussian random
variables:

\begin{equation}
        \kappa_{\alpha}\, =\, \frac{1}{\sqrt{N(N+1)}} \sum_{i<j}^{N} \tilde{J}_{ij} \alpha_i \alpha_j\, , \qquad \qquad
        V_{\alpha} \, =\, \frac{1}{\sqrt{N+1}} \sum_{i=1}^N J_{i,N+1} \alpha_i \, .
\end{equation}

Thus, we get
\begin{equation}\label{eq:cavity_step}
\E\Big(\ln\, \frac{Z_{N+1}}{Z_N}\Big)\, =\, \E\bigg(\ln\, \frac{\sum_{\alpha,\hat{\sigma}} \xi_{\alpha} e^{\beta (V_{\alpha} \, \hat{\sigma} + h \hat{\sigma})}}{\sum_{\alpha} \xi_{\alpha} e^{\beta \kappa_{\alpha}}}\bigg)\, ,
\end{equation}
where $\{\xi_{\alpha}\}$ are the weights
\begin{equation}
        \xi_{\alpha}\, =\, \exp\Big(\beta\Big[\frac{1}{\sqrt{N+1}} \sum_{i<j}^{N} J_{ij} \alpha_i \alpha_j
+ \underline{h} \cdot \underline{\alpha} \Big]\Big)\, .
\end{equation}

Equation~\eqref{eq:cavity_step} expresses the incremental contribution to the free energy in terms of the mean free energy of a particle added to a reservoir whose internal state is described by $\alpha$, corrected by an inverse-fugacity term ($\kappa$).  The latter may be thought of as the free energy of a `place holder', or a vacancy, for the cavity into which the $(N+1)$st particle is added.

\subsection{The cavity dynamics}  \label{cavitydyn}

One may note that the addition of a particle to the reservoir of $N$ particles has an effect on the state of the reservoir.   For $N>>1$, the value of the added spin, $\hat{\sigma}$, does not affect significantly  the field which would exist for the next increment in $N$, the direct contribution being only of the order $O(1/\sqrt{N})$. Hence,   for the next addition of a particle we may continue to regard  the state of the reservoir as given by just the configuration $\alpha$.  However, the weight of the configuration  (which is still to be normalized to yield its probability) undergoes the change:
\begin{equation}
  \xi_{\alpha} \mapsto    \xi_{\alpha}\, \sum_{\hat{\sigma} = \pm 1} e^{\beta (V_{\alpha} \, \hat{\sigma} + h \hat{\sigma})} \, .
\end{equation}
We refer to this transformation of the state of the reservoir (i.e., its probability distribution) as the {\em cavity dynamics}.

\subsection{Random Overlap Structures}

The state of the reservoir is relevant in so far as it correlates
with the cavity field $V_{\alpha} $ and fugacity  variable
$\kappa_{\alpha} $. In order to keep track of just the relevant
information, it is natural to introduce the following concept of a
{\em random overlap structure}~\cite{AS2}.    The definition is
somewhat tentative, as we do not address here the possibility that
the a continuum of states will be needed  for the reservoir, in
the limit $N\to \infty$.  (One may envision an extension of the
definition, but that will require addressing some technical
issues.)   Instead, we consider the case that states of the
reservoir form just a countable collection, which we order by the
weights. Even this simple concept allows to formulate variational
bounds, and in fact even capture Parisi's ansatz.

\begin{defn} A random overlap structure (ROSt) is a probability space $(\Omega, \mu)$ over which there are defined:
{\em i.\/}  a monotone nondecreasing sequence $\{\xi_n(\omega)\}$, and {\em ii.\/} an $\N \times \N$  matrix $\{q_{n,n'}(\omega)\}$ such that for $\mu$ a.e. $\omega$
\begin{itemize}
\item[(1)]  $\xi_n(\omega) \geq 0$ and $0<\sum_n \xi(\omega)_n<\infty$;
\item[(2)] $q_{n,n'}(\omega)$ corresponds to a real, positive semidefinite form;
\item[(3)] $q_{n,n}(\omega)=1$ for all $n\in \N$ (which implies $|q_{n,n'}|\leq 1$).
\end{itemize}
\end{defn}

Here, for clarity of the concept, we label the states of the reservoir not by  $\alpha$, as above, but by $n\in \N$.  However, as we shall see below, in the presence of the additional structure the somewhat vaguer notation will be convenient.   We shall not change the symbol for the weights  but rather just tacitly assume  that the sequence $\{ \xi_n\}$ is ordered, whereas $\{ \xi_{\alpha}\}$ is just a collection of the weights attached to an index which may have some additional structure, as will be encountered below.

\subsection{The incremental free energy functional}

In the above discussion, we presented the cavity dynamics as the
process of adding a single spin. But one can also add directly $M$
spins.  To describe the effect of that, one may associate with
each state of the ROSt new independent families of centered
Gaussian random variables$\{ \eta_{\alpha}^i \}_{i=1,\dots,M}$ and
$ \kappa_{\alpha} $  with the covariances
\begin{equation}   \label{eq:eta_cov}
\E(\eta_{\alpha}^i \eta_{\alpha'}^j)\, =\, \delta_{i,j} \, q_{\alpha,\alpha'}   \,
\quad ,  \qquad
\E(\kappa_{\alpha} \kappa_{\alpha'})\, =\, \frac{1}{2} q_{\alpha,\alpha'}^2\,   .
\end{equation}
For the added $M$-spin configuration $\{ \sigma_i\}_{1=1 \dots M}$ we define
\begin{equation} \label{eq:vas}
V_{\alpha,\sigma}\, =\, \sum_{i=1}^M \eta_{\alpha}^i \sigma_i\, .
\end{equation}

Motivated by the above consideration of the incremental free
energy in case the ROSt is just the SK system of $N$ particle, we
define the more general ROSt  functional:
\begin{equation} \label{eq:mfef}
        G_M(\mu)\, \stackrel{{\rm def}}{=}\, \frac{1}{M} \E\bigg(\ln\, \frac{\sum_{\alpha,\sigma} \xi_{\alpha} e^{\beta(V_{\alpha,\sigma} + \underline{h}\cdot \underline{\sigma})}}{\sum_{\alpha} \xi_{\alpha} e^{\beta \sqrt{M}\kappa_{\alpha}}}\bigg)\, .
\end{equation}
To ensure that this functional is well defined, let us note:
\begin{lem} \label{lem:integrability} For any configuration of the ROSt,
\begin{equation} \label{finite}
\ln\bigg(\frac{\sum_{\alpha} \xi_{\alpha} e^{\beta \sqrt{M}\kappa_{\alpha}}}{\sum_{\alpha} \xi_{\alpha}}\bigg)\,   \quad \mbox{and} \quad
\ln\bigg(\frac{\sum_{\alpha,\sigma} \xi_{\alpha} e^{\beta(V_{\alpha,\sigma} + \underline{h}\cdot \underline{\sigma})}}{\sum_{\alpha} \xi_{\alpha}}\bigg)\,
\end{equation}
are integrable with respect to the Gaussian measure averages over $\kappa_{\alpha}$ and $V_{\alpha,\sigma}$ (denoted below by $\E_{\kappa,V}$).
\end{lem}
\begin{proof}  To estimate the mean of the absolute value, it is convenient to use the identity,
$|X| = X + 2|-X|_{+} $,  for  $X\in R$.
Applying the Jensen's inequality to the average over $\alpha$ we get
 \begin{equation}
   - \ln\, \frac{\sum_{\alpha} \xi_{\alpha} e^{\beta\sqrt{M} \kappa_{\alpha}}}{\sum_{\alpha} \xi_{\alpha}}  \ \le \
 \frac{\sum_{\alpha} \xi_{\alpha} \ \beta\sqrt{M} |\kappa_{\alpha}|} {\sum_{\alpha} \xi_{\alpha}}
\end{equation}
With another application of the  Jensen inequality, this time to the average over the Gaussian variables, $\E_{\kappa} ( \ln Q) \le \ln \E_{\kappa} ( Q)$,  we get
 \begin{eqnarray}
\E_{\kappa} \bigg(\bigg|\ln\, \frac{\sum_{\alpha} \xi_{\alpha} e^{\beta\sqrt{M} \kappa_{\alpha}}}{\sum_{\alpha} \xi_{\alpha}}\bigg|\bigg) &
\leq &
\ln \frac{\sum_{\alpha} \xi_{\alpha} \E(e^{\beta\sqrt{M} \kappa_{\alpha} }) } {\sum_{\alpha} \xi_{\alpha}}\
+ \  2  \frac{\sum_{\alpha} \xi_{\alpha} \ \beta\sqrt{M} \E_{\kappa} (|\kappa_{\alpha}|)} {\sum_{\alpha} \xi_{\alpha}}
\nonumber \\
& \le & \frac{\beta^2 M}{4} +  \beta \sqrt{2 M} \
<\ \infty\ .
\end{eqnarray}
Similar bounds apply to the second quantity in \eqref{finite}.
\end{proof}

It should be clear from the above discussion and some elementary
estimates, as the one given below,  that in case the ROSt is just
the system of $N>>M$ particles with the Gibbs equilibrium state
corresponding to the SK interaction ($\mu_N$),
\begin{equation}
G_M(\mu_N)\ = \   \frac{1}{M}\E( \log [ Z_{N+M}
(\beta,h)/Z_N(\beta,h) ]  ) \ + \ O(1/M) \,  .
\end{equation}
However, rather surprisingly, it turns out that quite generally
the ROSt functional provides an {\em upper bound}:  

\begin{thm}[AS$^2$, a generalization of Guerra's  bound~\cite{Gue} ]
\label{ROStbound} For any ROSt:
\begin{equation}
\frac{1}{M} \E(\ln Z_M)\, \leq\, G_M(\mu) + o(1)\, ,
\end{equation}
where $o(1)$ vanishes for $M\to \infty$.
\end{thm}
Furthermore, one gets the following expression for the difference:
\begin{equation}
G_M(\mu) - \frac{1}{M}\E(\ln Z_M)\, =\, \frac{\beta^2}{2} \int_0^1 dt\, \E_t^{(2)}((q_{\sigma,\sigma'} - q_{\alpha,\alpha'})^2)\, .
\end{equation}
Here
$\E_t^{(2)}( \cdot )$ is a double replica average which is defined in
Section~\ref{subsec:gvp}, where this proposition is proved as part of Theorem~\ref{thm:as2}.

\noindent {\bf Remark:\/}
\noindent  {\bf 1.}  There is an interesting similarity, but also contrast on which we comment next, between Theorem~\ref{ROStbound} and the {\em Gibbs variational principle}.   For an arbitrary Hamiltonian $H(\sigma)$, and the initial probability measure $\rho_{0}(d\, \sigma)$, any probability distribution on the spins, $\mu(d\, \sigma)$, yields a variational {\em lower bound} for the logarithm of the partition function $ Z\, =\, \sum_{\sigma} e^{\beta H(\sigma)}$:
\begin{equation}\label{eq:Gibbs}
\ln\, Z\, \geq\, S(\mu|\rho_0) - \beta\, \mu(H)\, ,
\end{equation}
where $S(\mu|\rho_0)$ is the relative entropy of $\mu$ with respect to $\rho_0$, and $\mu(H)$ is the expectation value of $H$ with respect to $\mu$.   The inequality is saturated (for a finite system) if and only if $\mu$ is the Gibbs equilibrium state
$\rho(\sigma)\, =\, \frac{e^{-\beta H(\sigma)}}{Z}\, \rho_0(\sigma)$.

\noindent   {\bf 2.}   It is thus curious that the ROSt
variational principle yields upper bounds on the quenched free
energy, whereas the usual Gibbs variational estimate yields lower
bounds. We owe to Anton Bovier the interesting observation that
this change may be related to one of the puzzles encountered in
Parisi's original argument.  There, in the replica calculation the
usual role of minima and maxima are reversed due to the change of
sign in $n(n-1)$ when $n\to 0$.

\noindent  {\bf 3.}  As would be explained below, restricting the
variational bound to the hierarchal ROSt, RPC, one obtains the
result of Guerra \cite{Gue} that Parisi's solution provides an
upper bound on the pressure (lower bound on the free energy).


\section{Interpolation arguments}\label{gen_var_prin}

\subsection{A Gaussian differentiation formula}

The derivation of the variational principle rests on the following
differentiation formula.

\begin{lem} \label{lem:dif} Let $\Gamma$ be a finite index set and
$\left\{ X_{\gamma} \right\}_{ \gamma \in \Gamma}$ be a sequence
of centered, gaussian random variables whose correlations depend on a parameter $t\in(0,1)$:
\begin{equation} \label{eq:xcor}
\E_t\left( X_{\gamma} X_{\gamma'} \right) \, = \, C_{\gamma,
\gamma'}(t) \, ,
\end{equation}
with $C_{\gamma, \gamma'}(t) $ differentiable in $t$ and uniformly positive ($C_{\gamma, \gamma'}(t) \ge \varepsilon {\mathbb I}$) as a quadratic form.

Then, for any function
$\psi : \mathbb{R}^{\dim \Gamma} \to \mathbb{R}$ with  continuous second
partial derivatives that are polynomially bounded:
\begin{equation} \label{eq:diflem}
\frac{d}{dt}  \E_t\left(  \psi \left( \{ X_{\gamma} \} \right)
\right) \, = \, \frac{1}{2} \sum_{\gamma, \gamma^{\prime}}
  \frac{d}{dt}C_{\gamma, \gamma^{\prime}}(t) \,
\E_t\left( \frac{ \partial^2 \psi}{ \partial X_{\gamma} \partial
X_{\gamma^{\prime}} } \right)\, .
\end{equation}
\end{lem}

For polynomial functions $\psi$ the differentiation formula can be
obtained rather directly from Wick's rule \cite{Simon}, or through
the  integration by parts formula for gaussian random variables.
In appendix~\ref{app:gaussian} we present a proof based on the
Fourier transform representation.  The statement can be further
extended to functions whose second derivatives increase slower
than any inverse gaussian.

In our applications, we will be differentiating functions
of a specific form.  For this reason, we state:

\begin{cor} \label{cor:dif} Let $\{ X_{\gamma} \}_{\gamma \in
\Gamma}$ be a collection of Gaussian random variables as in
Lemma~\ref{lem:dif}, with
\begin{equation} \label{reg}
\sup_t \sup_{\gamma,\gamma' \in \Gamma} \left|\frac{d}{dt}
C_{\gamma,\gamma'}(t)\right|<\infty\, ,
\end{equation}
and $  \{ \xi_{\gamma} \}_{\gamma \in \Gamma} $
 a summable sequence of positive numbers.
Let
\begin{equation} \label{eq:psi}
\psi( \{ X_{\gamma} \}) := \ln \bigg(\sum_{\gamma \in \Gamma}
\xi_{\gamma} e^{ -\beta X_{\gamma}} \bigg),
\end{equation}
with some  $\beta > 0$.  Then, for any $0 \le t_1 < t_2 \le 1$
\\
\begin{eqnarray} \label{eq:cordif}
 \E_{t_2}\left( \psi \left( \{ X_{\gamma} \} \right) \right)
-  \E_{t_1}\left( \psi \left( \{ X_{\gamma} \} \right) \right) \,
    = \,   \qquad  \qquad   \\[2ex]
 =   \frac{\beta^2}{2} \int_{t_1}^{t_2}
 \left[
   \E_t^{(1)}\left(\frac{d}{dt} C_{\gamma, \gamma}(t) \right)
-  \E_t^{(2)}\left( \frac{d}{dt} C_{ \gamma, \gamma'} (t)
\right)\right]\, dt \, ,  \nonumber
\end{eqnarray}
where $\E_t^{(n)}$ represent the  ``weighted replica averages'',
which are defined, for bounded functions $f: \Gamma^n \to \R$, by
\begin{equation} \label{eq:repav}
\E_{t}^{(n)} \left( f( \gamma_1, \ldots, \gamma_n) \right) := \Ev{
\sum_{\gamma_1, \ldots, \gamma_n} f( \gamma_1, \ldots, \gamma_n)
  \prod_{i=1}^n \zeta_t(\gamma_i ) },
\end{equation}
with
\begin{equation} \label{eq:wts}
\zeta_t( \gamma) := \frac{ \xi_{\gamma} e^{- \beta X_{\gamma}}}{
\sum_{\gamma^{\prime}} \xi_{\gamma^{\prime}} e^{- \beta
X_{\gamma^{\prime}}}}.
\end{equation}
\end{cor}

\begin{proof}
For $\Gamma$ a finite set, the statement is a direct application
of \eqref{eq:diflem}.  For infinite
$\Gamma=\{\gamma_1,\gamma_2,\dots\}$ let $\Gamma_n =
\{\gamma_1,\dots,\gamma_n\}$. Then, as just stated,
\begin{equation} \label{eq:cordifFINITE}
\frac{d}{dt} \E\left( \ln \sum_{\gamma \in \Gamma_n} \xi_{\gamma}
e^{-\beta X_{\gamma}}\right)\, =\, \frac{\beta^2}{2} \left[
\E_{n,t}^{(1)}\left(\frac{d}{dt} C_{\gamma, \gamma}(t) \right) -
\E_{n,t}^{(2)}\left( \frac{d}{dt} C_{ \gamma, \gamma'} (t)
\right)\right]\, ,
\end{equation}
where the annealed multi-replica measures $\E_{n,t}^{(1)}$ and
$\E_{n,t}^{(2)}$ are with respect to the random discrete measure
generated by the finite sequence
\begin{equation} \label{eq:wtsFINITE}
\zeta_{n,t}( \gamma) := \frac{ \xi_{\gamma} e^{- \beta
X_{\gamma}}}{ \sum_{\gamma^{\prime} \in \Gamma_n}
\xi_{\gamma^{\prime}} e^{- \beta X_{\gamma^{\prime}}}}\, =\,
\frac{\zeta_t(\gamma)}{ \sum_{\gamma^{\prime} \in \Gamma_n}
\zeta_t(\gamma^{\prime})}\, ,
\end{equation}
for $\gamma \in \Gamma_n$. Thus, the statement holds for the
finite subsets $\Gamma_n$.  As $n \to \infty$ the random measures
determined by $\zeta_{n,t}$ converge to the random measure
determined by $\zeta_t$, e.g., in the total variation norm. The
claimed \eqref{eq:cordif} then follows using the integrated
version of \eqref{eq:cordifFINITE}, \eqref{reg}, and the bounded
convergence theorem.
\end{proof}

\noindent {\bf Remarks: 1. \/}
 The subscript $t$ above indicates that these averages depend on
the external parameter $t$ through the weights (\ref{eq:wts}).

{\bf 2. \/}  The derivative of $\psi$ separates into two crucial terms.
In many applications, the term involving the single replica
average, $\E_t^{(1)}( \cdot)$, will vanish because the variance of
$X_{\gamma}$ (i.e., the diagonal term) will remain constant with respect to the
interpolation parameter $t$.  For such cases, one sees that if the off diagonal terms
$C_{\gamma, \gamma'}(t)$  only decrease with $t$, then the function
$\Ev{\psi}$ increases in $t$.  Stated differently: the average
goes up when the variables $X_{\gamma}$ become  less
correlated.

  {\bf 3. \/}    Lemma~\ref{lem:dif} and Corollary~\ref{cor:dif} are
related to Slepian's inequality, c.f. \cite{Joagdev}.

While various interesting conclusions follow from monotonicity alone, it helps to go beyond that.  Following is a useful bound.

\begin{Corollary} \label{cor:est2} Suppose $\{ X_{\gamma} \}$ and $\{ Y_{\gamma} \}$
are two independent sequences of centered gaussian random
variables. Suppose that $\psi$ is as in Corollary \ref{cor:dif}. Then
\begin{equation} \label{eq:corest}
\big| \E \big( \psi(\{X_\gamma\})\big) - \E\big(\psi(\{Y_\gamma\})\big)\big|\, \leq\, \beta^2 \max_{\gamma,\gamma'} \big|
  \Ev{ X_{\gamma} X_{\gamma'}} - \Ev{Y_{\gamma} Y_{\gamma'}} \big|
\end{equation}
\end{Corollary}
\begin{proof}
Consider the Gaussian family $\{Z_{\gamma}\}$ with covariance
\begin{equation}
C_{\gamma,\gamma'}(t)\, =\, t\,  \E(X_{\gamma} X_{\gamma'}) +
(1-t)\,  \E(Y_{\gamma} Y_{\gamma'})\, .
\end{equation}
By Corollary \ref{cor:dif} one obtains a formula for the derivative of $\E\big(\psi(\{Z_{\gamma}\})\big)$, which can be bounded by the right-hand-side of (\ref{eq:corest}) at every $t\in(0,1)$.
\end{proof}
We note that if  the variances of $X_{\gamma}$ and
$Y_{\gamma}$ are equal for each $\gamma$ then $\beta^2$ can be replaced  by $\beta^2/2$ in (\ref{eq:corest}).


\subsection{GT interpolation and sub-additivity of the free energy}
The gaussian differentiation formula
(\ref{eq:diflem}) permits a quick derivation of the fundamental result of
Guerra and Toninelli \cite{GT} proving the existence of the
free energy for the SK model.

In order to state their result it is useful to include
extra diagonal terms in the Hamiltonian. These have a vanishingly
small effect in the $N\to\infty$ limit, but allow for the simplest
statement of the theorem.
\begin{equation} \label{eq:skhamd}
-H_N(\sigma)\, =\, \frac{1}{\sqrt{2N}} \sum_{i,j=1}^{N} J_{ij}
\sigma_i \sigma_j + \underline{h} \cdot \underline{\alpha} \, ,
\end{equation}
where the $J_{ij}$ are i.i.d.\ $N(0,1)$ random variables.  This
changes the covariance matrix entries by an amount of order
$O(1/N)$. Therefore, by Corollary \ref{cor:est2} it does not
affect $P(\beta,h)$. Henceforth, all $P_N(\beta,h)$, etc., are
defined relative to this Hamiltonian.  Alternatively, one can
define a centered Gaussian process $\{K_N(\sigma)\, , \sigma \in
\{+1,-1\}^N\}$ with covariance
\begin{equation} \label{eq:Kdef}
\E(K_N(\sigma)K_N(\sigma'))\, =\, \frac{N}{2} q_{\sigma,\sigma'}^2\, ,
\end{equation}
and then $-H_N(\sigma)\, =\, K_N(\sigma) + \underline{h} \cdot \underline{\sigma}$.

The first application of the interpolation is the super-additivity
of the quenched free energy
\begin{equation} \label{eq:qp}
Q_N(\beta, h) := \Ev{ \ln \left[ Z_N( \beta,h) \right] } \, .
\end{equation}
\begin{thm}(Guerra-Toninelli\cite{GT}) \label{thm:gt} For any
 $N, M \in \N$,
\begin{equation} \label{eq:qpsa}
Q_N( \beta, h) + Q_M( \beta, h) \leq Q_{N+M}( \beta,h).
\end{equation}
\end{thm}

\begin{proof}  Consider a system of size
$N+M$ with spin configurations $\gamma = \{ \gamma_i \}_{i=1}^{N+M}$.
Write the configuration as $\gamma = ( \alpha,\sigma)$,
with $\alpha = \{ \alpha_i \}_{i=1}^{N} := \{ \gamma_i \}_{i=1}^{N}$
and $\sigma = \{ \sigma_i \}_{i=1}^{M}
:= \{ \gamma_{N+i} \}_{i=1}^{M}$.  The Hamiltonian
for the  system of $N+M$ spins is given by
\begin{equation}
H_{N+M}( \gamma, h) = - K_{N+M}(\gamma)-
 \underline{h} \cdot \underline{\gamma} \, .
\end{equation}
Guerra and Toninelli have noted the utility of considering the one-parameter family of Hamiltonians which interpolate between $H_{N+M}$ and the sum of two independent SK Hamiltonians:
 \begin{equation} \label{eq:inths}
H_{N+M}( \gamma, h;t) := - K_{N+M}( \gamma;t) -  \underline{h}
\cdot \underline{\gamma} \, ,
\end{equation}
with
\begin{equation} \label{eq:ints}
K_{N+M}( \gamma; t) := \sqrt{1-t} \left[ K_N( \alpha) + K_M( \sigma) \right] + \sqrt{t} K_{N+M}( \gamma)\, .
\end{equation}
It is to be understood here that the random variables defining the interaction terms $K_N(\alpha)$, $K_M(\sigma)$, and
$K_{N+M}$, as in (\ref{eq:Kdef}),
are each chosen independently.  The function
\begin{equation}
\psi(t) := \ln \left[ \sum_{\gamma} e^{ - \beta H_{N+M}( \gamma, h;t)} \right],
\end{equation}
clearly satisfies
\begin{equation}
\begin{array}{ccc}
\Ev{ \psi(0) } = Q_N( \beta, h) + Q_M( \beta,h) & \mbox{and} &
\Ev{\psi(1)} = Q_{N+M}( \beta, h), \end{array}
\end{equation}
where $\Ev{ \cdot}$ above stands for integration with respect to all
the random couplings in $K_N$, $K_M$, and $K_{N+M}$. The theorem now follows
if we can control the sign of $\frac{d}{dt}\Ev{ \psi(t)}$.

To apply our differentiation formula,  (\ref{eq:cordif}), we
note that
\begin{equation} \label{eq:dc}
 \frac{d}{dt} C_{ \gamma, \gamma'} (t)  = \frac{N+M}{2} q_{\gamma,\gamma'}^2 - \frac{N}{2} q_{\alpha, \alpha'}^2 -
\frac{M}{2} q_{\sigma, \sigma'}^2.
\end{equation}
For the diagonal terms, $\gamma = \gamma'$,   the
above derivative vanishes.    Moreover, $q_{\gamma, \gamma'}$ is a convex
combination of $q_{\alpha, \alpha'}$ and $q_{\sigma, \sigma'}$:
\begin{equation} \label{eq:decqtt}
q_{\gamma, \gamma'} =  \frac{N}{N+M} q_{\alpha, \alpha'} +
\frac{M}{N+M} q_{\sigma, \sigma'}.
\end{equation}
Convexity of the function $f(q) = q^2$, allows to conclude that
\begin{equation} \label{eq:dcneg}
 \frac{d}{dt} C_{ \gamma, \gamma'} (t) \leq 0,
\end{equation}
and therefore, $\Ev{ \psi(t) }$ is increasing by (\ref{eq:cordif}).
This completes the proof.  
\end{proof}

Theorem \ref{thm:gt} immediately implies the existence of the
thermodynamic limit:
\begin{cor} \label{cor:pex} i) For any $\beta$ and $h$.
\begin{equation} \label{eq:avpe}
P( \beta, h) := \lim_{N \to \infty} P_N(\beta,h)\,
\end{equation}
exists.
Furthermore, defining $\mathcal{P}_N(\beta,h;\omega) = \frac{1}{N} \ln\, Z_N(\beta,h;\omega)$,
\begin{equation} \label{eq:pe}
\lim_{N \to \infty} \mathcal{P}_N( \beta, h;\omega) = P( \beta, h),
\end{equation}
where the limit is in distribution.
\newline \noindent ii) The pressure may also be represented as
\begin{equation} \label{eq:sap}
P( \beta, h)  = \lim_{M \to \infty} \liminf_{N \to \infty} \frac{1}{M} \Ev{ \ln
  \left[ \frac{Z_{N+M}( \beta, h)}{ Z_N( \beta, h)} \right]},
\end{equation}
for any $\beta$ and $h$.
\end{cor}

It should be  noted here that prior to the GT argument it was known that the fluctuations of  $\mathcal{P}_N(\beta,h;\omega)$ are of diminishing size as $N\to \infty$, a fact which can be deduced by either martingale methods \cite{PS} or a concentration of measure argument \cite{Tal}.    The `monotonicity of the interpolation' argument \cite{GT} adds the last missing step, which is the convergence of the sequence  $P_N(\beta,h)$.

{\it Proof of Corollary~\ref{cor:pex}:}
The results claimed in
(\ref{eq:avpe}) and (\ref{eq:sap}) are simple consequences of
(\ref{eq:qpsa}); namely if $\{ Q_N \}_{N \in \mathbb{N}}$ is a super-additive
sequence, then the following limit exists and may be calculated
incrementally
\begin{equation} \label{eq:limqn}
\lim_{N \to \infty} \frac{Q_N}{N} = \lim_{M \to \infty} \liminf_{N \to
  \infty} \frac{Q_{N+M}- Q_N}{M},
\end{equation}
see Lemma~\ref{lem:charlim} below.

The convergence  \eqref{eq:pe}  follows from  (\ref{eq:avpe})
since, as, mentioned above, the range of the probability
distribution of $\frac{1}{N} \ln\, Z_N(\beta,h;\omega)$ narrows as
$N\to \infty$  - a fact proven in \cite{PS,Tal}. \hfill \qed

\vspace{.3cm}

\noindent {\bf Remarks: 1. \/}
While (i) recovers the Guerra and Toninelli result \cite{GT},
(ii) is an observation which was useful in the proof of the
variational principle \cite{AS2}.

\noindent {\bf 2. \/} The reader is cautioned that the
super-additivity of the quenched pressure, and the particular
direction for its monotonicity under the process of `amalgamation'
in which two blocks are interpolated into a single system, is not
a thermodynamic principle akin to the Gibbs-phenomenon.  For the
Curie-Weiss model the inequality in (\ref{eq:qpsa}) is reversed
(as  a simple calculation will show).


\section{The ROSt variational principle} \label{subsec:gvp}

For convenience, let us remind ourselves that the functional representing the increase in the free energy due to the incorporation of $M$ spins $\sigma \in \{+1,-1\}^M$ into a  ROSt $\mu$ whose configurations are described by $(\{\xi_{\alpha}\}_{\alpha}, \{q_{\alpha,\alpha'}\})$ is
\begin{equation}\label{eq:Gdef}
        G_M(\beta,h;\mu)\, =\, \frac{1}{M} \E\bigg(\ln\, \frac{\sum_{\alpha,\sigma} \xi_
        {\alpha} e^{\beta(V_{\alpha,\sigma} +  \underline{h} \cdot \underline{\gamma})}}
        {\sum_{\alpha} \xi_{\alpha}
        e^{\beta\sqrt{M}\kappa_{\alpha}}}\bigg)\, .
\end{equation}
with $\{ \kappa_{\alpha}, V_{\alpha,\sigma} \} $ Gaussian random variables of covariance,
\begin{align} \label{eq:kappa_cov2}
\E(\kappa_{\alpha} \kappa_{\alpha'})\, &=\, \frac{1}{2} q_{\alpha,\alpha'}^2\, ,\\
\label{eq:V_cov}
\E(V_{\alpha,\sigma} V_{\alpha',\sigma'})  &=\, M q_{\alpha,\alpha'} q_{\sigma,\sigma'}\, .
\end{align}

\begin{thm} \label{thm:as2}
{\it 1.} For any $N$ and any ROSt $\mu$
\begin{equation}\label{eq:GBd}
\frac{1}{N}\, \E(\ln\, Z_N(\beta,h))\, \leq\, G_N(\beta,h;\mu)
\end{equation}
with
\begin{equation}\label{eq:GDiff}
G_N(\beta,h;\mu) - \frac{1}{N} \E(\ln Z_N(\beta,h))\,
=\, \frac{\beta^2}{2} \int_0^1 dt\, \E_t^{(2)}((q_{\alpha,\alpha'}-q_{\sigma,\sigma'})^2)\, .
\end{equation}
{\it 2.} The pressure  is given by
\begin{equation}  \label{AS2}
P(\beta,h)\, =\, \lim_{M\to\infty} \inf_{\mu{\rm : ROSt}} G_M(\beta,h; \mu)\, ,
\end{equation}
where the limit $M\to\infty$ also equals  the supremum over $M$.
\end{thm}

The replica expectation $\E_t^{(2)}$ is just as in Corollary
\ref{cor:dif} with respect to the interpolating Gaussian process
with covariance
\begin{equation}
\E(X_{\gamma} X_{\gamma'})\,
=\, C_{\gamma,\gamma'}(t)\,
=\, \frac{N}{2} [(1-t) (q_{\alpha,\alpha'}^2 + q_{\sigma,\sigma'}^2) + 2  t\,  q_{\sigma,\sigma'} \, q_{\alpha,\alpha'}]\, ,
\end{equation}
where $\gamma=(\alpha, \sigma )$.

\begin{proof}
\noindent { \em  Part 1.:}
The argument is  a slight modification of the interpolation scheme described in
Theorem~\ref{thm:gt}.   Here we consider a system composed of a finite
block of spins $\sigma$, whose interactions are determined by the SK
model, and a reservoir of configurations $\alpha$, whose overlaps
are governed by a ROSt $\mu$. Again, we interpolate between a
decoupled state of the system and a state in which some interactions are allowed.
 The interpolating Hamiltonian is
\begin{equation} \label{eq:inth}
-H_N( \alpha, \sigma;t) := \sqrt{1-t} \left[ K_N( \sigma) + \sqrt{N}
  \kappa_{\alpha} \right]  + \sqrt{t} V_{\alpha, \sigma} +
   \underline{h} \cdot \underline{\gamma} \, ,
\end{equation}
where the random couplings in $K_N( \sigma)$, $\kappa_{\alpha}$,
and $V_{\alpha, \sigma}$, defined in (\ref{eq:Kdef}) (\ref{eq:kappa_cov2}), and (\ref{eq:V_cov})
respectively, are each drawn independently.

The function
\begin{equation} \label{eq:psit}
\tilde{\psi}(t) := \frac{1}{N} \ln \left[ \frac{\sum_{\alpha, \sigma}
    \xi_{\alpha} e^{- \beta H_N( \alpha, \sigma; t)} }{ \sum_{\alpha}
    \xi_{\alpha} e^{ \beta \sqrt{N} \kappa_{\alpha}}} \right],
\end{equation}
is easily seen to satisfy
\begin{equation} \label{okbcs}
\Ev{ \tilde{\psi}(0)} = P_N( \beta, h) \ \ \mbox{and} \ \ \Ev{ \tilde{
    \psi}(1)} = G_N( \beta, h; \mu),
\end{equation}
where $\Ev{ \cdot }$ stands for integration with respect to all
random variables appearing in (\ref{eq:psit}).

Our differentiation formula (\ref{eq:cordif}) applies again.
Letting $\gamma$ now denote pairs $\gamma = ( \alpha, \sigma)$, we have $-H_N(\gamma;t) = X_\gamma + \underline{h} \cdot \underline{\sigma}$ where a
direct calculation shows that
\begin{eqnarray} \label{dcor2}
\frac{d}{dt} C_{ \gamma, \gamma'} (t)  =
- \frac{N}{2} \left( q_{\alpha, \alpha'} - q_{\sigma, \sigma'} \right)^2\, .
\end{eqnarray}
The covariance derivative vanishes for $\gamma = \gamma'$, since $q_{\alpha, \alpha} =  q_{\sigma, \sigma}=1$;  as we saw, already that and the definite sign in \eqref{dcor2} imply monotonicity.   The full statement,
(\ref{eq:GDiff}), follows by (\ref{eq:cordif}) and the fundamental theorem of calculus.

\noindent {\em  Part 2.:}
We now note
that there exists a sequence of ROSts $\mu_N^{SK}$ for which
\begin{equation} \label{eq:sat}
 G_M( \beta, h; \mu_N^{SK}) =   \frac{1}{M} \Ev{ \ln
  \left[ \frac{Z_{N+M}( \beta, h)}{ Z_N( \beta, h)} \right]} \ + \
  o(\frac{M}{N}) \, .
\end{equation}
To see that, it suffices to consider the example which has
motivated the concept, namely the case when the  ROSt,
$\mu_N^{SK}$ is provided by another SK systems of $N$ particles,
with $N>> M$.

Adapting \eqref{eq:cavity_step} to an increment by $M$ we get
\begin{equation}\label{eq:cavity_M}
\E\Big(\ln\, \frac{Z_{N+1}}{Z_N}\Big)\, = \, \E\bigg(\ln\,
\frac{\sum_{\alpha, {\sigma}} \xi_{\alpha} e^{\beta ( V_{\alpha,
\sigma} + \underline{h}\cdot \underline{\sigma}) )}}{\sum_{\alpha}
\xi_{\alpha} e^{\beta \kappa_{\alpha}}}\bigg)\, ,
\end{equation}
where $\{\xi_{\alpha}\}$ are the weights
\begin{equation}
        \xi_{\alpha}\, =\, \exp\Big(\beta\Big[\frac{1}{\sqrt{N+M}}
        \sum_{i<j}^{N} J_{ij} \alpha_i \alpha_j
+ \underline{h} \cdot \underline{\alpha} \Big]\Big)\, .
\end{equation}
The quantities $V_{\alpha,\sigma}$ and $\kappa_{\alpha}$  are
Gaussian variables whose covariance differs from the corresponding
factors in the desired variational quantity by the factor of $
\frac{N}{N+M} = 1-\frac{M}{N} + O(\, (\frac{M}{N} )^2) $. Applying
 Corollary~\ref{cor:est2}, one may determine that
\begin{equation}\label{eq:Gbd}
 \left| G_M( \beta, h; \mu^{SK}_N) - \frac{1}{M} \Ev{ \ln \left(
      \frac{Z_{N+M}( \beta, h)}{ Z_N( \beta,h)} \right) } \right|
= O \left( \frac{M}{N}   \right) \, ,
\end{equation}
from which (\ref{eq:sat}) follows.
 
Combining this result with Corollary~\ref{cor:pex} part (ii) gives
part (2) of Theorem \ref{thm:as2}.
\end{proof}


\section{Hierarchal Random Probability Cascades (RPC)} \label{sec:RPC}

In his commentary on the story of Oedipus, Andre Gide brought up the observation that there exist universally valid answers, which are applicable to many questions.
~\footnote{In the case of Oedipus, ``I/man'', points towards the answer to the two questions which Oedipus faced at turning points in his life, the one posed by the Sphinx and the other on which years later he has sought the advice of Tiresias. {\it A. Gide: ``Oedipe'' (1931).}  } 
A ``universal answer'', in the form of a hierarchal structure which appears to play a key role in various complex systems, has emerged also in the study of spin-glass models.  

In this section, we describe a family of ROSts each of which is endowed with a remarkable property: quasi-stationarity under a class of time evolutions which includes the  cavity dynamics of Section~\eqref{cavitydyn}.
An intriguing and relevant question is
whether the class of examples discussed here includes
all the ROSts which exhibit a robust version of
quasi-stationarity.  Before explaining the question, or conjecture,
let us present the ``random energy model'' and its hierarchal
extension.   Both were introduced by Ruelle, as the point processes
capturing the $N\to \infty$ limit of Derrida's finite model
calculations, and called the REM (for random
energy model) and the GREM  (for a generalized random energy
model).   Seeking a descriptive term we shall refer to these
as the hierarchal ``random probability cascades'' (RPC).

\subsection{The Random Energy Model (REM)} \label{subsec:REM}

The basic building block for the hierarchal probability cascades
is the REM, or REM$_x$ to be specific, which is the Poisson point process on
$[0,\infty)$ with density given by  $\rho_x(d \xi) = - d\,  \xi ^{-x}$.
Here $x$ is a parameter ranging over $ (0,1)$, the minus sign is to
ensure that the measure is positive, and each configuration,
drawn according to the REM$_x$, is represented by a sequence
of non-negative numbers denoted by $\{ \xi_{\alpha} (\omega)\}$.
Denoting the occupation number
of a Borel set $A \subset [0, \infty)$ by
\begin{equation}
N_A( \omega) := \# \{ \alpha : \xi_{\alpha}( \omega) \in A \},
\end{equation}
what is stated above means that for the REM$_x$: \\
 {\em i.}  the occupation numbers of disjoint sets form independent random variables, \\
  {\em ii.}  the distribution of the
occupation number is Poissonian:
\begin{equation} \label{int}
\mathbb{P} \left( N_A( \omega) = k \right) = \frac{ \rho(A)^k}{k!}
\, e^{- \rho(A)} \,  ,
\end{equation}
with $\rho(A)$ the mean value:
\begin{equation} \label{eq:ocrho}
 \rho_x(A)\ \equiv \ \Ev{N_A( \omega)} \  =
 \   - \int_A d \,  \xi ^{-x}  \,  .
\end{equation}

The REM$_x$ process also appears in extreme value theory; in some
probability references it is denoted $\textrm{PD}(x,0)$,  (\cite{Pitman1997}.)

By (\ref{eq:ocrho}), for any $\epsilon >0$
\begin{equation} \label{eq:expo}
   \E\left(  N_{[\epsilon, \infty)}( \omega) \right) \, = \, \frac{1}{\epsilon^x} \, .
\end{equation}
It readily follows that with probability one
 it is possible to re-label its points in descending
order, i.e., write $\{ \xi_{\alpha} ( \omega) \} = 
\{ \xi_n ( \omega) \}_{n=1}^{\infty}$
where
\begin{equation} \label{eq:order}
 \xi_1( \omega) > \xi_2( \omega) > \cdots \, .
\end{equation}
Furthermore, one has:
\begin{thm} \label{prop:rem} Let $0<x<1$, 
then with respect to the point process REM$_x$, almost surely: \\
 \mbox{i.     }
 \begin{equation} \label{n_xi}
 \qquad   n^{1/x} \xi_n(\omega)  \, \too{n\to \infty} \, 1 \, .
\end{equation}
 \mbox{ii. the following sum converges if and only if $v >x$:} \qquad
\begin{equation} \label{norm}
  \sum_{n} \xi_{n} ( \omega)^v \, <  \, \infty \, ,
\end{equation}
 \mbox{iii.  } the partition function
$  Z(\omega) := \sum_{n} \xi_{n} ( \omega)   $ is almost surely finite,
with an infinitely divisible distribution, satisfying the addition law: $Z\, \stackrel{\mathcal{D}}{=}\, 2^{-1/x}(Z + Z')$ where $Z'$ is an iid copy of $Z$, \\
 \mbox{iv. the $u$-moment of $Z$ is finite if and only if
 $u < x$:} \qquad
 \begin{equation} \label{Z_u}
   \E \left( Z^u \right) \,  < \, \infty \, .
 \end{equation}
 where $  \E$ represents the expectation value over REM$_{x}$.
\end{thm}

\begin{proof}

{\em i.\/} On the scale of  $t\equiv \xi^{-x}$, REM$_x$ is a
Poisson process of fixed density $(=1)$. Let $N(t;\omega) := N_{(
t^{- \frac{1}{x}}, \infty)}( \omega)$, and let $t(n;\omega) $ be
the inverse function.  Then, by the Law of Large Numbers (or the
ergodic theorem), $N(t; \omega) /t \to 1$, almost surely (for
$t\to \infty$).   This can be rewritten as $t(n;\omega) /n\to 1$,
which implies \eqref{n_xi}. (Estimates on the deviations can be
deduced using the law of the iterated logarithm.)

{\em ii.\/}  The a.s. finiteness statement \eqref{norm} can be
deduced from \eqref{n_xi}, or alternatively by splitting from the
sum the finite (almost surely) collection of terms with $\xi >1$,
and noting that the main term is then of finite mean.

{\em iii.\/}  The divisibility law for the distribution of
$Z(\omega)$ is a direct consequence of the divisibility of the
Poisson point process.

{\em iv.\/}  A simple device which facilitates the derivation of
\eqref{Z_u} is the bound, for $0\le u \le 1$:
\begin{equation}
Z(\omega)^u \ \le \ Z_{[0,1]}(\omega)^u+\sum_{\xi_n \in  (1,\infty)} \xi_n^u(\omega)
\end{equation}
where  $Z_{[0,1]}$ is the contribution due to $\xi_n \in  (0,1]$. With the help of  the H\"older inequality, at $p=1/u$, applied to the first term,  \eqref{Z_u}  can be deduced by a direct calculation.
 \end{proof}


\subsubsection{Quasi-Stationarity of the REM} \label{subsubsec:qs}

Among the more compelling attributes of the REM point processes,
is their {\it quasi-stationarity} under the  dynamics which
correspond to increments through independent factors.

The time evolution can be described through a sequence of steps
applied to a configuration $\{ \xi_{\alpha} (\omega)\}$ generated according
to a REM$_x$ process.  First, the points of the configuration
are labeled in descending order
$\{ \xi_{\alpha} \} = \{ \xi_n \}_{n=1}^{\infty}$ as described in \eqref{eq:order}.
Here we have omitted the dependence of the sequence on the randomness
$\omega$, and we will continue to do so, where convenient, in the
following.
Next, a non-negative sequence of iid random  variables
$\{ \gamma_n \}_{n=1}^{ \infty}$
is drawn independently of $\{\xi_n \}$,
with probability distribution $g(d \gamma)$.
A new configuration is obtained by multiplying $\xi_n$ by the random weights
$\gamma_n$.  To retain the monotonicity which is assumed in our notation, the resulting configuration
is relabeled in descending order, and it therefore takes the form
\begin{equation}
\widetilde{\xi}_n := \gamma_{\pi(n)} \xi_{\pi(n)}
\end{equation}
where $\pi$ is the appropriate permutation.
We also denote
\begin{equation}
\widetilde{\gamma}_n := \gamma_{\pi(n)} \, .
\end{equation}
Thus, while $\gamma_n$ is the factor by which $\xi_n$ is multiplied
``going forward'' in time,  $\widetilde{\gamma}_n$ is the
factor by which $\widetilde{\xi}_n$ was increased in the last step.

\begin{thm} \label{thm:qs}  For any $x \in (0,1) $ and a
probability   distribution $g(d \gamma) $ of finite moment:
  $ \langle \gamma^x \rangle := \int \gamma^x\, g(d \gamma) < \infty $, there is a constant $K$ so that the REM$_x$ distribution is stationary under the
 time evolution produced by the random factors $\{ \gamma_n \}$, as described above, corrected by the factor
 \begin{equation}  \label{K}
 K \ = \ \langle \gamma^x \rangle^{1/x} \, ,
 \end{equation}
in the sense that:
  \begin{equation} \label{quasi}
\{ K^{-1}\, \tilde{ \xi}_n \} \  \stackrel{ \mathcal{D}}{=} \
\{ \xi_n \}.
\end{equation}
Furthermore, the past increments $\{ \tilde {\gamma}_{n}( \omega) \}$
form a sequence of
iid  random variables with the modified probability distribution
\begin{equation} \label{wdist}
\widetilde{g}(d \widetilde{\gamma}) =
\frac{\widetilde {\gamma}^{x}\,  g(d \widetilde \gamma) }{ \langle \gamma^x     \rangle  }
\end{equation}
which are also  independent of  $\{   \widetilde \xi_n \}$.
\end{thm}

The last statement may appear paradoxical:  you start with a sequence
of the iid  random variables $\{ \gamma_n \}$,  reshuffle them a bit,
producing the permuted sequence $\{ \widetilde \gamma_n \}$, and the
result is a sequence of iid variables with a different distribution!
This would certainly not be possible for any finite collection of
random variables, but it is apparently possible in the infinite setting
due to the existence of a bottomless reservoir.

\begin{proof}[Theorem \ref{thm:qs}]  The proof  can be obtained  through the moment generating functionals, or alternatively the observation that 
the joint distribution of the collection
$\{( \xi_n, \gamma_n ) \}$
corresponds to the Poisson process in $\R_+ \times \R_+$ with the
density: $-d\, \xi^{-x} \, g(d \gamma)$.   The collection of points
 $\{(\widetilde{\xi}_n,\widetilde{\gamma}_n)\}$ also forms a Poisson process,
 since its occupation numbers for disjoint regions of $\R_+\times \R_+$
 are independent, and have the density
 $-d\, (\widetilde {\xi}/\widetilde {\gamma}) ^{-x} \, g(d \widetilde {\gamma}) $.   
 It helps to write this density so
 that it becomes a probability measure in the second variable: \begin{equation} \label{jdist}
- d\, (\widetilde {\xi}/\widetilde {\gamma}) ^{-x} \, g(d \widetilde {\gamma})
\, = \,
-d\, \left( \frac{\widetilde {\xi} } { \langle \gamma ^x
     \rangle^{1/x}}  \right )^{-x}    \times \frac{\widetilde {\gamma}^{x}\,  
     g(d \widetilde \gamma) }{ \langle \gamma^x \rangle  } \,  .
\end{equation}
The fact that the second factor on the right hand side is a probability measure which does not depend on $\widetilde \xi$
allows to quickly read from the above the statements which are asserted in the Theorem.
\end{proof} 

The above argument  is discussed a bit more explicitly in \cite{RA}.
Theorem~\ref{thm:qs} states that each of the  REM$_x$  processes is
invariant under the stochastic evolution up to a deterministic correction.
A general result of Liggett~\cite{Liggett} implies that such
invariance in fact singles out this class of processes.
A strengthening of this statement was obtained in the work of
Ruzmaikina and Aizenman~\cite{RA}.
For our applications, it suffices to know that the distribution
of the {\em relative weights} is stationary, in the sense that:
\begin{equation} \label{eq:qs}
\left\{ \frac{\xi_n }{Z} \right\} \,
\stackrel{ {\mathcal D}}{=}  \,
\left\{ \frac{\widetilde{\xi}_n}{ \widetilde{Z}}
\right\},
\end{equation}
where the partition function $Z$, resp. $\widetilde{Z}$, are as
introduced in Theorem~\ref{prop:rem} (iii). In ref.~\cite{RA} this
property was termed ``quasi-stationarity'', and  it was shown there
that, under certain limitations on the point process and the
distribution of the independent weights, just this property limits
the point process to REM$_x$ at some value of the parameter $x \in (0,1)$.


\subsection{The Random Probability Cascades (RPC)} \label{subsec:RPC}

The REM point processes were used by D. Ruelle as building blocks
for a hierarchal process which capture the results of Derrida's
calculations involving the large $N$ limit of the free energy in
the so-called Generalized Random Energy Models. In line with
Parisi's fundamental insight concerning the SK spin-glass model,
the parameter for the construction is a monotone function $x(q)$
taking $[0,1]$ into itself. Convenient examples, and
approximations, are provided by piecewise constant functions.  For
each $k \in \N$, a piecewise constant {\em right-continuous}
 function $x(q)$ is specified by a
pair of monotone sequences
\begin{eqnarray} \label{eq:xqpart}
0= x_0 < x_1 < x_2 < \cdots < x_k < x_{k+1} = 1 \,  , \nonumber \\
0= q_0 < q_1 < q_2 < \cdots < q_k < q_{k+1} = 1 \, ,
\end{eqnarray}
in particular,
\begin{equation} \label{eq:xofq1}
 x(q) := \sum_{i=0}^k x_i \chi_{[q_i, q_{i+1})}(q)
\end{equation}
with $x(1)=1$.

Following is the hierarchal construction parametrized by this
data. \\
i. Start with a REM$_{x_1}$ process whose points
are symbolically labeled as $\{ \xi^{(1)}_{\alpha_1} \}$.  Here the
subscript $\alpha_1$  is intended to represent a label which
just identifies the points; not their respective ordering.
(If absolutely desired, $\alpha_1$ could be regarded as
taking values in a random subset of the line.) \\
ii.  Next, for each $\alpha_1$, we generate a REM$_{x_2}$ process
whose points are designated $\xi^{(2)}_{\alpha_2;\alpha_1}$.  The processes
corresponding to different values of $\alpha_1$ are choosen independently. \\
iii.  The  construction is iterated up to $n=k$.  At the $n$-th step,
independent versions of the REM$_{x_n}$  process are generated
for each of the distinct values of the
``address'' $(\alpha_1, \alpha_2, ... , \alpha_{n-1})$,
and the resulting points are designated  as
$\xi^{(n)}_{\alpha_n;\alpha_1,..., \alpha_{n-1}}$.

The construction yields a hierarchal family of addresses of
the form
\begin{equation}
\underline{\alpha} \ =\   (\alpha_1,..., \alpha_{k}) \, .
\end{equation}
With each value of $\underline\alpha$, we associate
\begin{equation} \label{def:RPCxi}
\xi_{\underline{\alpha}} \ := \   \prod_{n=1}^k
\xi^{(n)}_{\alpha_n;\alpha_1,..., \alpha_{n-1}}  \, .
\end{equation}

The result of the above construction is a point process whose
configurations consist of the collection
 $   \xi(\omega) := \{ \xi_{\underline \alpha} \} $, where $\omega$
 - which is omitted on the right hand side- represents all
 the randomness which enters the above construction.
 (Specifically, all the above choices can be represented by functions
 defined over a probability space whose points are denoted
 by $\omega$.)

The hierarchal addresses, which play a role in the explicit
construction, can ipso-facto be  replaced by the more generic ROSt notation,
for which the information is expressed through the overlap
kernel, which here is defined as:
\begin{equation} \label{eq:haol}
q_{\alpha,   \alpha '} \ \equiv \  q_{n(\alpha,   \alpha ')} \, ,
\quad n(\alpha,   \alpha ') \ := \ \max \{j : j \le k \, , \,
(\alpha_1,..., \alpha_j) = (\alpha'_1,..., \alpha'_j)  \} \, .
\end{equation}
An overlap kernel corresponds to a hierarchal address if and only
if the condition:  $q_{ \alpha,  \alpha '} \ \le \  r$ is transitive for each real $r$.  The condition can equivalently be expressed as ``ultrametricity'' \cite{MPV} of the distance function $\dist( \alpha,  \alpha ') := 1- q_{ \alpha,  \alpha '} $.

Let us note that for $\mathbb{P}^{(2)}$ -- the probability measure associated to $\E^{(2)}$ -- a calculation yields \cite{Ru}
$$
\mathbb{P}^{(2)}(q_{\alpha,\alpha'} \geq q)\, =\, x(q)\, .
$$

As a direct consequence of Theorem~\ref{prop:rem} and Theorem~\ref{thm:qs} one has:

\begin{thm} \label{thm:RPCnorm} For $k \ge 2$, and $0<x_1 < ... < x_k <1$, the partition function
$Z = \sum_{\underline \alpha = (\alpha_1, ..., \alpha_k)} \xi_{\underline \alpha} \quad $
is almost surely finite and in distribution satisfies:
\begin{equation} \label{eq:normprod}
Z\  \stackrel{ {\mathcal D} }{=} \   Z_{x_1} \, \prod_{n=2}^{k}
\E\left(  [Z_{x_{n}}]^{x_{n-1} }  \right)^{ 1/x_{n-1}}
\end{equation}
where $ Z_{x_{n}}$ is a random variable having the distribution of a partition function  under REM$_{x_n}$.
   In particular,
\begin{equation} \label{eq:logz}
 \E ( \log Z ) < \infty \, .
\end{equation}
\end{thm}

The above construction yields a process whose configurations consist of
the pair:
\begin{equation}
( \{  \xi_{\underline \alpha}(\omega) \}_{\underline \alpha},
\,  \{ q_{\underline \alpha, \underline \alpha '}(\omega) \}_{\underline \alpha, \underline \alpha'} )
\end{equation}
of: i.  a point subset of $[0,\infty)$, and ii. an overlap kernel,
 which conveys the genealogical information.   Our main interest
 will concern the system of normalized weights, along with the
 overlaps, i.e., 
\begin{equation} \label{normRPC}
( \{  \{  \xi_{  \alpha}(\omega) / Z(\omega) \} ,
\,  \{ q_{\underline \alpha, \underline \alpha '}(\omega) \}_{  \alpha,   \alpha'} )
\end{equation}
 We refer to this process  as the Random Probability Cascade.

 \noindent {\bf Remark:}  The last step in the hierarchal construction should correspond to REM$_{x_{k+1}}$ at $x_{k+1}=1$, which may be seen as problematic since for $x= 1$ the normalization $Z$ diverges.  Nevertheless,  for $x\nearrow 1$, the normalized average is well defined for all the quantities of interest.  For simplicity of the presentation we shall not stress this  point here, and approach the value $x=1$  only as a limit.

\subsection{Quasi-stationarity of  RPC}

The hierarchal  RPC inherits and broadens the remarkable
quasi-stationarity property of the REM processes. In the context
of RPC, the dynamics allow also correlated evolution of the point
configuration. The construction of the evolution is similar to
that considered for the REM model, except that the random factors
$\gamma_n$ are now of the form:
\begin{equation}
\gamma_n \ = \ e^{\psi(\eta_n ) }
\end{equation}
with $\{\eta_n\}$  a collection of Gaussian random
variables of covariance
\begin{equation} \label{eq:etacov}
\E ( \eta_{n} \eta_ {n'}   ) \  = \ q_{\alpha(n), \alpha(n')} \, ,
\end{equation}
where $\alpha=\alpha(n)$ is the inverse of the bijection $n=n(\alpha)$.

Unlike the previous case, the dynamics are now correlated. The
correlations between the increments  of the ``competing'' points
are determined through the overlap function, but are not affected
by the relative ranking of their position on the line, which
changes in the course of the time evolution.

It is important to note that the covariance condition
\eqref{eq:etacov} is satisfiable, i.e., the hierarchal kernel
$q_{n(\underline \alpha), n(\underline \alpha' )} $ is always
positive definite. To see that, it is useful to construct an
auxiliary genealogical tree for which the ultrametric kernel
coincides with the value of $q$  at which the ancestral lines of
$\underline \alpha$ and $\underline \alpha' $ split. A Gaussian
process with the covariance \eqref{eq:etacov} is obtained by
associating with each $\underline \alpha$ the integral of white
noise along the branches of the tree, in a path leading from the
root to $\underline \alpha$, with the covariance
$\E\left((d\eta)^2 \right)=dq $.
Furthermore, by restricting the white noise
integral to only $q\in [0,t]$, one obtains a family of Gaussian
variables with an extra parameter $t\in [0,1]$, $\eta_n(t) $, with
the covariance:
\begin{equation} \label{eq:tcov}
\E ( \eta_n(t) \eta_ {n' }(t)   ) \  = \ \min \left
\{ t, q_{n(\underline \alpha), n(\underline \alpha' )} \right \} \, .
\end{equation}

A convenient explicit representation is obtained by presenting  the Gaussian variables
$\eta_{n(\underline \alpha)}$ as sums of mutually independent
terms, which in the  algorithm described above correspond to the
integrals of white noise over distinct segment of the genealogical
tree:
\begin{equation} \label{decomp}
  \eta_{ \underline \alpha }\ \equiv \ \eta_{n(\underline \alpha)}\
 = \ \sum_{i=1}^k
\sqrt{q_{i+1} - q_i} \, Z_{i, \underline{\alpha}}\, ,
\end{equation}
where $Z_{i, \underline{\alpha}}$ are normal Gaussian variables
with the covariance
\begin{equation}
\E (Z_{i, \underline{\alpha}} Z_{j, \underline{\alpha'}})\ = \
\delta_{i,j} \, I[q_{ \underline \alpha ,  \underline \alpha' }
\ge q_{j+1} ] \, .
\end{equation}

For a simple statement of the quasi-stationarity, a relevant class
of function is defined by the Lipschitz norm:
\begin{equation}
\|\psi  \|_{Lip} \ := \ \sup_{x,y} \frac{| \psi(x) - \psi(y) |} {|x-y|}
\end{equation}

\begin{thm} \label{thm:grem}
Under the dynamics described above, for any $\psi$ of bounded
Lipschitz norm, the configuration which results from the above
dynamics has the same distribution as the process obtained by
multiplying $\{ \xi_{\underline \alpha} \}$ by a constant, $
e^{\psi_0 } $:
\begin{equation} \label{eq:rpcpsiqs}
\{  \widetilde \xi_{\underline \alpha}(\omega) \} \ :=
\ \{  e^{\psi(\eta_{\underline \alpha}(\omega)) }
 \xi_{\underline \alpha}(\omega)  \} \
\stackrel{ {\mathcal D}}{=} \
\{  e^{\psi_0 }   \xi_{\underline \alpha}(\omega)  \} \, ,
\end{equation}
with $\psi_0$ described below.
In particular, the partition function satisfies
\begin{equation}
\widetilde Z \stackrel{ {\mathcal D} }{=}   e^{\psi_0 } Z \, .
\end{equation}
and the process is quasi-stationary, in the sense that the
distribution of the relative weights
$\{  \xi_{\underline \alpha} (\omega) / Z(\omega) \} $
is stationary, satisfying the appropriate version of eq.~\eqref{eq:qs}.
\end{thm}

\begin{proof} This statement can be obtained by a direct iteration of the
quasi-stationarity property of the REM processes which are used in
the construction of the RPC.
It is convenient to define the partial quantities, for any $j =1, \ldots, k$:
 \begin{equation} \label{xij}
\xi_{\underline{\alpha}}^{(j)} \ :=\  \prod_{n=1}^j \xi_{\alpha_n;
\alpha_1,
  \dots, \alpha_{n-1}}^{(n)} \quad \quad \mbox{and}
\quad \quad \eta_{\underline{\alpha}}^{(j)}\ :=\  \sum_{i=0}^j
\sqrt{q_{i+1} - q_i} \, Z_{i, \underline{\alpha}} \, .
\end{equation}

Conditioning on the collection of variables
$\eta_{\underline{\alpha}}^{(k-1)}$ and
$\xi_{\underline{\alpha}}^{(k-1)}$ and for each
$\alpha_1,...,\alpha_{k-1}$ let us consider the evolution for the
corresponding subtree which corresponds to multiplication by
\begin{equation} \label{eq:wks}
\gamma_{k,\underline{\alpha}}\ := \ e^{ \psi \left(
\eta_{\underline{\alpha}}^{k-1} + \sqrt{q_{k+1} - q_k} \, Z_{k,
\underline{\alpha}} \right)} \, \xi_{\underline{\alpha}}^{k-1} \,
.
\end{equation}
By  Theorem \ref{thm:qs}, for each  subfamily corresponding to a
specified $\alpha_1,...,\alpha_{k-1}$:
\begin{equation} \label{wkpush}
\left\{ e^{ \psi( \eta_{\underline{\alpha}})} \,
  \xi_{\underline{\alpha}} \right\}  = \left\{
\gamma_k \, \xi_{\alpha_k ; \alpha_1, \dots, \alpha_{k-1}}
\right\}  \stackrel{ \mathcal{D}}{=} \, \left\{ \langle
\gamma_k^{x_k} \rangle ^{1/x_k} \, \, \xi_{ \alpha_k ; \alpha_1,
\dots, \alpha_{k-1}} \right\},
\end{equation}
where $\langle \cdot \rangle$ represents integration with respect
to the variables $Z_{k, \underline{\alpha}}$.

 The above procedure of conditioning and averaging may be
iterated. Starting from: $ \psi_k(y) \equiv \psi $, and denoting
by $\E_z ( \cdot )$  the average over the normal gaussian random
variable $z$, we define recursively for $j: \ k\searrow 0$:
\begin{equation}
\psi_j(y) := \frac{1}{x_j} \ln \left[ \E_z \left( e^{ x_j \cdot
\psi_{j+1}
      \left( y + \sqrt{q_{j+1} - q_j} \cdot z \right) } \right) \right]
      \, ,
\end{equation}
It is easy to check that under the Lipschitz condition on $\psi$
 the iteration step is well defined, and, furthermore, the
Lipschitz norm does not increase under the mapping $\psi_j(\cdot)
\mapsto \psi_{j-1}(\cdot) $. One obtains
\begin{equation}  \label{psi0}
\left\{ e^{ \psi( \eta_{\underline{\alpha}})} \,
\xi_{\underline{\alpha}} \right\}
\  \stackrel{  \mathcal{D}}{=} \
  \dots  \left\{ e^{ \psi_1 \left(
      \eta_{\underline{\alpha}}^{{0}} \right)} \,
  \xi_{\underline{\alpha}} \right\}
  \  \stackrel{  \mathcal{D}}{=} \
  \left\{ e^{ \psi_0 }\,
  \xi_{\underline{\alpha}} \right\}
  \, .
\end{equation}
In this sequence, the deterministic quantity appearing in
(\ref{eq:rpcpsiqs}) is
\begin{equation}
\psi_0 := \ln \left[ \E_z \left( e^{\psi_1 \left( \sqrt{q_1} z
\right)} \right) \right] \, .
\end{equation}
\end{proof}

The deterministic value of $\psi_0$ can be alternatively
characterized through the solution of a specific partial
differential equation.  Namely, consider functions of two
variables $f=f(q,y)$ which satisfy, for $t\in [0,1]$
\begin{equation} \label{eq:pde0}
\frac{ \partial f}{ \partial q} + \frac{1}{2} \left[ \frac{
\partial^2
  f}{ \partial y^2} + x(q) \left( \frac{\partial f}{ \partial y} \right)^2 \right] = 0,
\end{equation}
with the $t=1$ boundary condition:
\begin{equation} \label{eq:bc0}
f(1,y) = \ln \left[ \cosh \left[ \beta ( y + h ) \right] \right].
\end{equation}
One may note that the function $x(q)$ enters here as a parameter
for the partial differential equation.  Going backward in time,
the equation is particularly simple to solve  over intervals where
where $x(q)$ is constant.  Using the Cole-Hopf transformation, on
which more is said next, the solution is provided by the iterative
  procedure which is described in the above proof.
From this perspective, the value of $\psi_0$ corresponds to
$\psi_0 \ = \ f(0,0;x)$.  We shall now expand on this point.

\subsection{Quasi-stationarity of RPC  in terms of the  Parisi equation}

An alternative perspective on Theorem~\ref{thm:grem} is provided
by a continuous time version of the quasi-stationarity.  As it
turns out,  equation \eqref{eq:pde0}, which plays a key role in the
Parisi solution, appears also as a Martingale condition for the
cavity dynamics with respect to the RPC hierarchal ROSt.    

For a given function $\psi$ consider the
two parameter function $f(t,y)=f(t,y;x)$, which satisfies the
boundary conditions:
\begin{equation} \label{eq:bc00}
f(1,y) = \psi( y + h ).
\end{equation}
and the partial differential equation \eqref{eq:pde0},
which is to be solved from $q=1$ \underline {down} to $q=0$.

Theorem \ref{thm:grem} admits the following extension, about 
which  we learned from D. Ruelle.   
For simplicity it is implicitly assumed here that the function is
suitably differentiable and bounded.  Upon closer analysis, it
suffices to assume the Lipschitz condition, as in
Theorem~\ref{thm:grem}.   

\begin{thm} \label{thm:t-grem}
Let $f(t,y)$ be a function satisfying
\eqref{eq:pde0}.   For configurations of  the hierarchal RPC which
correspond to a piecewise constant function $x(q)$, let:   
\begin{equation}
\xi_{\underline \alpha}(t; \omega) \ := \ e^{f(t,\eta(t;\omega))}
\, \xi_{\underline \alpha}(\omega) \, .
\end{equation}
Then the probability  distribution of the ROST configuration
$\xi_{\underline \alpha}(t; \omega)$ is independent of $t$.  In
particular, it coincides with that of $e^{\psi_0} \,
 \xi_{\underline \alpha}(\omega)$
where the deterministic factor is
\begin{equation}
 \psi_0 \ = \  f(0,0)
\end{equation}
\end{thm}

The statement can be proved along the lines of the above proof of
Theorem~\ref{thm:grem}, or in terms of stochastic PDE and Ito's
formula. We refer the reader for further details on the latter
perspective to \cite{Arg}.

Over intervals of constant $x(q)$ the differential equation can be
solved through the convolution of the function
$e^{f(q,\eta)/x(q)}$ with suitable Gaussian measures. This is a
slight variation of the well-known Cole-Hopf transform familiar in
the context of nonlinear integrable PDE's.

In the special case of  $x(\cdot)$  constant over the entire
interval $(0,1)$  the RPC is really a $\textrm{REM}_x$. In this
situation,  one readily verifies that the solution of
\eqref{eq:pde0}  derived through the Cole-Hopf transformation,
starting with the boundary conditions (\ref{eq:bc00}), at $ t=1$,
is exactly what one would obtain using \eqref{quasi}. For
piecewise constant $x(q)$ this argument can be employed in steps,
to again conclude that the PDE formulation matches with the
results of an iteration of  Theorem \ref{thm:qs}, i.e.,
\eqref{psi0}.  Subdividing the intervals into short segments the
statement  can also be easily understood from the perspective  of
Ito's formula, as is discussed more explicitly in \cite{Arg}.

The formulation of the
 solution in terms of the differential equation has the advantage
of being well defined even when the piecewise constant $x(q)$ is
replaced by a continuous function.   For the existence of the 
continuum limit it is imperative to restrict the attention to the 
ROSt given by the normalized weights, as  in \eqref{normRPC}.   

 Let us now return to the spin glass model for  whose solution the above
 plays a key role.


\section{Relation with the Parisi solution}

\subsection{The Parisi formula}

The partial differential equation, \eqref{eq:pde0}
has made its appearance  in the work of Parisi on the SK model, in
the context of rather different considerations. Without reviewing here Parisi's approach, and his hierarchal ansatz for replica symmetry breaking, let us present the resulting conjecture for the free energy, a.k.a. the `Parisi solution'.

Introducing the ansatz of hierarchal pattern of replica symmetry breaking - a concept for which the reader is referred to \cite{P,MPV} - Parisi has introduced the idea that the order parameter for the SK model is a monotone function, $x: \, [0,1] \mapsto [0,1]$.   Somewhat analogously to the much simpler case of the Curie Weiss mean field ferromagnetic model, the value of the order parameter can be characterized through either self consistency, based on  the cavity analysis of the cavity dynamics (discussed in Chapters 4 and 5 of \cite{MPV}), or through a variational principle.   That has led Parisi   to investigate solutions $f=f(q,y)$ of the partial differential equation
\begin{equation} \label{eq:pde}
\frac{ \partial f}{ \partial q} + \frac{1}{2} \left[ \frac{
\partial^2
  f}{ \partial y^2} + x(q) \left( \frac{\partial f}{ \partial y} \right)^2 \right] = 0 \,  ,
\end{equation}
subject to the boundary condition
\begin{equation} \label{eq:bc}
f(1,y) = \ln \left[ \cosh \left[ \beta ( y + h ) \right] \right] \,  .
\end{equation}
The resulting value of   $f(0,0)\equiv  f(0,0;x)$ is incorporated in the Parisi  functional, which is defined as:
\begin{equation} \label{eq:pfun}
P[x] \, := \, \ln[2] \, + \, f(0,0;x) - \, \frac{
  \beta^2}{2} \int_0^1 \, q \, x(q) \, dq \, .
  \end{equation}
The end result is Parisi's proposal that:
\begin{equation}
\lim_{N\to \infty}  \frac{1}{N} \log \E(Z_N)) \  =
\ \inf_{x(\cdot)} P[x]    \  := \   G_{Parisi} \, .
\end{equation}
where the infimum is over monotone functions of the unit interval with values in $[0,1]$.

The remarkable arguments of Parisi are still beyond mathematical analysis, but its main conclusion is now known to be   correct.

In a surprising development, F. Guerra~\cite{Gue} proved:
\begin{lem}
[Guerra variational principle]
\begin{equation}
\ \inf_{x(\cdot)} P[x]    \  \le  \   G_{Parisi} \, .
\end{equation}
\end{lem}
The analysis, which employs an  interpolation argument, yields also a criterion for the saturation of the inequality.
The statement was given a different form in our work~\cite{AS2}:  the variational principle was generalized into infimum of the functional $G(\beta,h;\mu)$ over ROSt's ($\mu$), and it was shown that
in that generality the infimum yields the correct value (Theorem~\ref{thm:as2}).
Independently of that, M. Talagrand~\cite{Tal_AM} has proven that the Parisi conjecture is correct. The proof employs the criterion provided by Guerra's analysis, and insights supported by a heavy dosage of calculus.

We shall now show how Guerra's variational principle is
incorporated in the ROSt bound, \eqref{eq:GBd} of
Theorem~\ref{thm:as2}.

\subsection{The free energy of Hierarchal ROSts}

For the hierarchal RPC, the calculation of the ROSt functional
$G_M(\beta,h;\mu) \equiv G_M(\mu)$ is greatly facilitated by their
quasi-stationarity property.   We shall now demonstrate that the
free energy functional corresponding to the RPC of  a given
function $x(q)$  is independent of $M$ and coincides with Parisi's
functional $P[x]$, of \eqref{eq:pfun}.

The ROSt   free energy functional, which is defined in
(\ref{eq:Gdef}), can be written  as
\begin{equation} \label{eq:g=g1-g2}
G_M( \mu) = G^{(1)}_M( \mu) - G^{(2)}_M( \mu)
\end{equation}
with
\begin{equation} \label{eq:g1}
G^{(1)}_M( \mu) = \frac{1}{M} \mathbb{E} \left( \ln \left[
    \frac{ \sum_{\alpha, \sigma} \xi_{\alpha}
e^{ \beta(V_{\alpha, \sigma} + \underline{h}\cdot
\underline{\sigma})} }{\sum_{\alpha} \xi_{\alpha} } \right]
\right)
\end{equation}
and
\begin{equation} \label{eq:g2}
G^{(2)}_M( \mu) = \frac{1}{M} \mathbb{E} \left(\ln \left[ \frac{
\sum_{\alpha, \sigma} \xi_{\alpha} e^{ \beta \sqrt{M}
\kappa_{\alpha} } } {\sum_{\alpha} \xi_{\alpha} } \right] \right).
\end{equation}

\begin{lem} \label{lem:rostfe}
Let $\mu$ be a ROSt having weights generated by an RPC with parameter
$x=(x_1,\dots,x_n)$ and overlap function $q$.  Then for any $M \in \mathbb{N}$:
\begin{equation} \label{eq:rostg1}
G^{(1)}_M( \mu) \, = \, \ln[2] \, +
\, f(0,0;x),
\end{equation}
and
\begin{equation} \label{eq:rostg2}
G^{(2)}_M( \mu ) \, = \,
\frac{\beta^2}{2} \, \int_0^1 \, q \, x(q) \, dq \, .
\end{equation}
In particular, the free energy functional coincides with the Parisi
functional at $x$, i.e.,
\begin{equation} \label{eq:g=a}
G_M(\beta,h;\mu) \, = \, P[x].
\end{equation}
\end{lem}

\begin{proof}
Summing over the spins $\sigma$, we cast $G^{(1)}_M$ in the form
\begin{equation} \label{eq:g1nos}
G^{(1)}_M( \mu )  \, =  \, \ln[2] \, + \, \frac{1}{M} \Ev{ \ln
\left[ \frac{ \sum_{\underline{\alpha}} \xi_{\underline{\alpha}}
 \prod_{i=1}^M e^{ \psi(\eta_{i, \underline{\alpha}})}
}{\sum_{\underline{\alpha}} \xi_{\underline{\alpha}} } \right] },
\end{equation}
where
\begin{equation} \label{eq:psieta}
\psi( \eta_{i,\underline{\alpha}}) \, := \, \ln [ \cosh[ \beta(
\eta_{i, \underline{\alpha}} +h)]] \, ,
\end{equation}
and $\eta_{i, \underline{\alpha}}$ are Gaussian variables with the covariance:
\begin{equation}
\E( \eta_{i, \underline{\alpha}}\, \eta_{i, \underline{\alpha}'}) \ = \
\delta_{i,i'}\,  q_{\underline{\alpha},\underline{\alpha}'}  \, .
\end{equation}
The  quasi-stationarity of the ROSt readily implies that the
contributions of the independent factors
$e^{\psi(\eta_{i,\underline{\alpha}})}$ of the right hand side in
\eqref{eq:g1nos} factorizes, and thus $G^{(1)}_M$ is independent
of $M$.  Furthermore,  by Theorem~\ref{thm:t-grem}, we see that
\begin{equation}
\left\{ \xi_{\underline{\alpha}} \prod_{i=1}^M e^{ \psi(\eta^i_{
      \underline{\alpha}})} \right\} \stackrel{\mathcal{D}}{=} \left\{
  e^{M f(0, 0;x)} \xi_{\underline{\alpha}} \right\} \, .
\end{equation}
This proves (\ref{eq:rostg1}).

We calculate $G^{(2)}_M$
by interpolation: For any $t \in [0,1]$, define
the function
\begin{equation}
F(t) := \frac{1}{M} \mathbb{E} \left(\ln \left[ \frac{
\sum_{\underline{\alpha}} \xi_{\underline{\alpha}} e^{ \beta
\sqrt{M} \sqrt{t} \kappa_{\underline{\alpha}} } }
{\sum_{\underline{\alpha}} \xi_{\underline{\alpha}} } \right]
\right).
\end{equation}
Note that
\begin{eqnarray}
F(1) = G^{(2)}_M(\mu ) & \mbox{and} &
F(0) = 0.
\end{eqnarray}
Using Lemma \ref{lem:dif}, we see that
\begin{eqnarray}
F'(t) = & \displaystyle \frac{\beta^2 }{2} \left( \frac{1}{2} -
\mathbb{E}^{(2)}_t \left(\frac{q_{\underline{\alpha},
\underline{\alpha}'}^2 }{2} \right) \right)  = & \displaystyle
\frac{\beta^2 }{2} \mathbb{E}^{(2)} \left(
\int_{q_{\underline{\alpha}, \underline{\alpha}'}}^1 \, q \, dq \right) \label{t=0} \\
\mbox{ } = & \displaystyle \frac{\beta^2 }{2} \int_0^1
\mathbb{P}^{(2)} \left( q_{\underline{\alpha}, \underline{\alpha}'}
\leq q \right) \, q \, dq = & \displaystyle \frac{\beta^2 }{2}
\int_0^1 x(q) \, q \, dq. \label{olnn}
\end{eqnarray}
In (\ref{t=0}) we have used quasi-stationarity of
$\xi_{\underline{\alpha}}$  to remove
the dependence on $t$.  Equation \eqref{eq:rostg2} follows through the integration of $F'(t)$.
\end{proof}

\subsection{An Open Problem: Explaining the validity of the
Parisi  ansatz}

As was mentioned above, it is now a Theorem, proven by M. Talagrand~\cite{Tal_AM}, that
Parisi's ansatz indeed yields the correct solution for the free
energy of the SK model.  However, it still seems reasonable to say
that an ``explanation'' of the reasons for the validity of the
Parisi ansatz continues to present an open challenge.
Could  RPC's be the only `robustly' quasi-stationary
ROSt's, and could the validity of Parisi's ansatz be explained by that?  Can one formulate some other fundamental reason for the validity of the Parisi calculation?  Given the versatility of the applications of the Parisi approach, it may be of interest to shed more light on any of  these questions.

\appendix

\section{The Gaussian Differentiation Lemma} \label{app:gaussian}

\begin{lem}
\label{lem:1} Let   $\X_t \in \R^n$, $t \in (0,1)$, be  a
vector-valued Gaussian process, with covariance $C_t$ which is
continuously differentiable. Suppose that $\psi:\R^n \to \R$ is
twice continuously differentiable and compactly supported. Then
\begin{equation}
\label{eq:1}
\frac{d}{dt} \E[\psi(\X_t)]\,
=\, \frac{1}{2} \E\left[\big(\ip{\nabla}{\dot{C}_t\nabla} \psi\big)(\X_t)\right]\, .
\end{equation}
\end{lem}

\begin{proof}
The joint density function for $\X_t$ is
\begin{equation}
\rho_t(\x)\, =\, \frac{\exp\left(-\frac{1}{2} \ip{\x}{C_t^{-1} \x}\right)}{\sqrt{\det(2\pi C_t)}}\, .
\end{equation}
In terms of the Fourier transform, $\hat{f}(\k)=\int_{\R^n}
e^{-2\pi i \ip{\k}{\x}} f(\x)\, d^n\x$,
\begin{equation}
\E[\psi(\X_t)]\, :=\, \int_{\R^n} \psi(\x)\, \rho_t(\x)\, d\x\,
=\, \int_{\R^n} \hat{\psi}(\k)\, \hat{\rho}_t(\k)\, d\k\, ,
\end{equation}
(by  Plancherel theorem).  Since $\hat{\rho_t}(\k)\, =\, \exp(-2
\pi^2 \ip{\k}{C_t \k})$, a direct calculation shows
\begin{equation}
\frac{d}{dt} \E[\psi(\X_t)]\,
=\, -2 \pi^2 \int_{\R^n} \ip{\k}{\dot{C}_t\k}\hat{\psi}(\k) \hat{\rho}_t(\k)\, d\k\, .
\end{equation}
But, since $(\nabla \psi)\, \hat{}\, (\k)\, =\, 2\pi i \k\, \hat{\psi}(\k)$, we see that
\begin{equation}
-2\pi^2 \ip{\k}{\dot{C}_t\k} \hat{\psi}(\k)\,
=\, \frac{1}{2} \big(\ip{\nabla}{\dot{C}_t\nabla} \psi\big)\,\hat{}\, (\k)\, .
\end{equation}
So, by Plancherel's theorem again,
\begin{equation}
\frac{d}{dt} \E[\psi(\X_t)]\,
=\, \frac{1}{2} \int_{\R^n} \big(\ip{\nabla}{\dot{C}_t\nabla} \psi\big)\,\hat{}\, (\k)\, \hat{\rho}_t(\k)\, d\k\,
=\, \frac{1}{2}  \E\left[\big( \ip{\nabla}{\dot{C}_t\nabla} \psi\big)(\X_t)\right]\, .
\end{equation}
\end{proof}

\smallskip
We need the following extension of this result to a wider class of
functions $\psi$, which is enabled by a density argument.

\begin{cor}
\label{cor:1}
Let $\X_t$  be as in Lemma \ref{lem:1}.
Suppose $\psi \in \mathcal{C}^2(\R^n)$
and $\psi,\nabla\psi,\nabla^2 \psi \in L^1(\R^n,\rho_t)$
for every $t \in (0,1)$.
Also suppose that
$$
\Big(t \mapsto \E\left[|\psi(\X_t)| + \|\nabla\psi(\X_t)\| + \|\nabla^2\psi(\X_t)\|\right] \Big)
\in L^1_{\text{\it loc}}((0,1))\, .
$$
Then $\E[\psi(\X_t)]$ is absolutely continuous and (\ref{eq:1}) holds for almost all $t \in (0,1)$.
\end{cor}

\begin{proof}
Let $\eta : \R^n \to \R$ be any smooth function, with compact support,
such that $0\leq \eta\leq 1$ and such that $\eta(0)=1$.
Define
$$
\psi_\epsilon(\x)\, =\, \eta(\epsilon \x) \psi(\x)\, ,
$$
for each $\epsilon>0$.
So $\psi_{\epsilon}$ is twice continuously differentiable, and with compact
support. Also, $\psi_{\epsilon} \to \psi$ and $\nabla^2 \psi_{\epsilon} \to \nabla^2 \psi$,
pointwise, as $\epsilon \to  0$.
Finally, we know that $|\psi_{\epsilon}(\x)| \leq |\psi(\x)|$ for all $\x \in \R^n$, and
\begin{equation}
\label{eq:silly}
\|\nabla^2 \psi_{\epsilon}(\x)\|\,
\leq K \left(|\psi(\x)| + \|\nabla \psi(\x)\| + \|\nabla^2 \psi(\x)\|\right)\, ,
\end{equation}
for some constant $K<\infty$. (The constant depends only on the
sup norm of $\|\nabla \eta(\x)\|$ and $\|\nabla^2 \eta(\x)\|$.)

By Lemma \ref{lem:1}, integrating,
$$
\E[\psi_{\epsilon}(\X_{t})] \Big|_{t_1}^{t_2}\,
=\, \frac{1}{2} \int_{t_1}^{t_2} \E\left[\big( \ip{\nabla}{\dot{C}_t\nabla} \psi_{\epsilon}\big)(\X_t)\right]\, dt\, ,
$$
for each $t_1,t_2 \in (0,1)$ and all $\epsilon>0$.
By the dominated convergence theorem,
$$
\lim_{\epsilon \downarrow 0} \E[\psi_{\epsilon}(\X_{t})]\, =\, \E[\psi(\X_{t})]
$$
for every $t \in (0,1)$. In particular, it is true at $t=t_1$ and $t=t_2$.
Similarly, by the dominated convergence theorem
$$
\lim_{\epsilon \downarrow 0}
\E\left[\big( \ip{\nabla}{\dot{C}_t\nabla} \psi_{\epsilon}\big)(\X_t)\right]\,
\, =\, \E\left[\big( \ip{\nabla}{\dot{C}_t\nabla}\psi\big)(\X_t)\right]\, ,
$$
for every $t \in [t_1,t_2]$.
But, moreover, the integral of the upper bound in (\ref{eq:silly}),
integrated against $\rho_t$,
is a function of $t$ which is locally integrable, by our hypothesis.
Therefore, we can apply the DCT to the $t$-integral, itself, to determine
$$
\lim_{\epsilon \downarrow 0}
\int_{t_1}^{t_2} \E\left[\big( \ip{\nabla}{\dot{C}_t\nabla} \psi_{\epsilon}\big)(\X_t)\right]\,
dt\, =\,
\int_{t_1}^{t_2} \E\left[\big( \ip{\nabla}{\dot{C}_t\nabla} \psi\big)(\X_t)\right]\, dt\, .
$$
So
$$
\E[\psi(\X_{t})] \Big|_{t_1}^{t_2}\,
=\, \frac{1}{2} \int_{t_1}^{t_2} \E\left[\big( \ip{\nabla}{\dot{C}_t\nabla} \psi\big)(\X_t)\right]\, dt\, .
$$
Since this is true for every $t_1,t_2 \in (0,1)$, Lebesgue's differentiation theorem
implies the corollary.
\end{proof}


\section{Limits for super-additive sequences}

In the proof of Theorem~\ref{thm:as2} we made use of the following
known statement.  For completeness we present its proof.
\begin{lem} \label{lem:charlim}
Let $\{ Q_N \}_{N \in \N}$ be a {\it super-additive} sequence of
real numbers, in the sense that for any $N, M \in \N$,
\begin{equation} \label{eq:add}
Q_N + Q_M \leq Q_{N+M}.
\end{equation}
Then the following limit exists, with value in $\R \cup \{ \infty \}$,  and satisfies
\begin{equation} \label{eq:lim1}
\lim_{N \to \infty} \frac{Q_N}{N} = \sup_{N} \frac{Q_N}{N} \, .
\end{equation}
Moreover,
\begin{equation} \label{eq:lim2}
\lim_{N \to \infty} \frac{Q_N}{N} = \lim_{M \to \infty} \liminf_{N \to
  \infty} \frac{Q_{N+M} - Q_N}{M}.
\end{equation}
\end{lem}

\begin{proof}
Let $M \in \N$. For any integer $N > M$, one may write
$N = n \cdot M + k$ with $1 \leq k < M$, and by super-additivity,
\begin{equation} \label{eq:lowbd2}
\liminf_{N \to \infty} \frac{Q_N}{N} \, \geq \, \frac{Q_M}{M}\, .
\end{equation}
Thus
\begin{equation} \label{eq:limex}
\limsup_{N \to \infty} \frac{Q_N}{N} \, \leq \, \sup_M \frac{Q_M}{M}
\, \leq \, \liminf_{N \to \infty} \frac{Q_N}{N},
\end{equation}
from which (\ref{eq:lim1}) follows.

The proof of (\ref{eq:lim2}) follows by demonstrating two
inequalities. An immediate consequence of (\ref{eq:add}), is
\begin{equation} \label{eq:upbd}
\frac{Q_M}{M} \, \leq \, \liminf_{N \to \infty} \frac{Q_{N+M} -Q_N}{M} \, ,
\end{equation}
and a lower bound, which is part of the claim  (\ref{eq:lim2}),
follows from the fact that the limit in (\ref{eq:lim1}) exists.

The matching upper bound may be obtained by noting that for any $n
\in \N$
\begin{equation} \label{eq:tel}
\frac{Q_{n \, M + N} - Q_N}{n \, M +N} \ \ = \ \ \frac{ \sum_{j=1}^{n}
  Q_{ j \, M+N} - Q_{(j-1) \, M + N}}{n \, M +N},
\end{equation}
and for each $j = 1, \ldots, n$
\begin{equation} \label{eq:increbd}
 Q_{j \, M+N} - Q_{(j-1) \, M + N} \, \geq \, \inf_{k \geq N} \left[
 Q_{k + M} - Q_k \right].
\end{equation}
Inserting (\ref{eq:increbd}) into (\ref{eq:tel}), taking $n \to \infty$,
and then the supremum over $N$, we arrive at
\begin{equation}
\lim_{n \to \infty} \frac{Q_n}{n} \geq \liminf_{N \to \infty}
\frac{Q_{N+M}- Q_N}{M}\,  ,
\end{equation}
which completes the proof of   (\ref{eq:lim2}).
\end{proof}

%
%

\section{General Interactions} \label{gen:int}

In this appendix, we will illustrate that the results provided in
the main text for the SK Hamiltonian have a simple analogue for
more general Hamiltonians. As was done in \cite{AS2}, we will
demonstrate that our analysis also holds for models of the type
\begin{equation} \label{eq:genham}
         H_N(\sigma, h) \, := \, - \, K_N(\sigma) \, - \,
         \underline{h} \cdot \underline{\sigma} \, ,
\end{equation}
where the interaction term $K_N( \sigma)$ is now taken to be a
centered Gaussian process, indexed by the spins $\sigma$, with the
covariance
\begin{equation} \label{eq:genKcov}
\E \left( K_N(\sigma) \,  K_N(\sigma') \, \right) \, = \,
\frac{N}{2} \, f( q_{\sigma,\sigma'}).
\end{equation}
Here, for convenience, $f$ is written as a function of the spin
overlap. We will assume that $f$ is a positive power series; i.e.,
$f(q):= \sum_{r=1}^{\infty}|a_r|^2 q^r$ on $[-1,1]$ with the
normalization
 $\sum_{r=1}^{\infty}|a_r|^2 =1$.
An explicit realization of such an interaction $K_N$ is given in
terms of the multi-spin interaction:
\begin{equation} \label{genint}
         K_N(\sig)\ = \ \sqrt{ \frac{N}{2}} \,
         \sum_{r=1}^\infty  \frac{a_r}{  N^{r/2}} \sum_{i_1,\dots,i_r=1}^N
         J_{i_1\dots i_r} \sigma_{i_1}\cdots\sigma_{i_r}
\end{equation}
where $J:= \{J_{i_1,\dots,i_r}\}$ is a family of independent
normal Gaussian variables.

For the results discussed here, we further assume that $f$ is
convex on $[-1,1]$. The importance of such a condition has been
recognized in the literature, e.g. in \cite{GT2} convexity was
used to prove  convergence for the free energy density, in the
limit $N \to \infty$. Derrida's $p$-spin models \cite{Der} are
obtained by the special choices $f(q) = q^p$ for $p \in \N$, and
for these convexity holds if $p \in 2\N$. In particular, setting
$p=2$, one recovers the SK model, except that in contrast to
(\ref{H}) the tensor in (\ref{genint}) need not be symmetric.  For
convenience we also include here diagonal terms, but these do not
affect the results.

In the analysis of the free energy it is convenient to first
assume that the second derivative of $f$ is continuous up to the
boundary, and then use continuity arguments for an extension of
the results.  We  proceed under this additional assumption.

The analogue of Corollary~\ref{cor:pex} and Theorem~\ref{thm:as2}
hold for Hamiltonians defined with the Gaussian interactions given
by (\ref{eq:genKcov}).
\begin{thm} \label{thm:AppB1} For any $\beta$ and $h$, define
$P_N( \beta, h)=\frac{1}{N} Q_N(\beta,h)$ relative the
Hamiltonian given by (\ref{eq:genham}). Then,
\begin{equation} \label{eq:genavpe}
P( \beta, h) := \lim_{N \to \infty}  P_N( \beta, h),
\end{equation}
exists, and moreover,
\begin{equation} \label{eq:genpe}
\lim_{N \to \infty} \mathcal{P}_N(\beta,h;\omega)\  =\
 P( \beta, h),
\end{equation}
almost surely.
\end{thm}

\begin{proof} With the very same interpolation scheme
(\ref{eq:inths}) and (\ref{eq:ints}), excepting that the random
variables are now defined via (\ref{eq:genKcov}), one derives
\begin{equation}
\frac{d}{dt} C_{\tau, \tau'}(t) \, = \, \frac{N+M}{2}f( q_{\tau,
  \tau'}) - \frac{N}{2}f( q_{\alpha, \alpha'}) \, - \, \frac{M}{2} f(
q_{ \sigma, \sigma'}),
\end{equation}
in place of (\ref{eq:dc}). Superadditivity, as before, follows
from the convexity $f$ and (\ref{eq:decqtt}).
\end{proof}

For the Hamiltonian given by (\ref{eq:genham}), one may also
develop a cavity perspective by performing the change in
free-energy analysis as described in Sections \ref{cav_perspect}
and \ref{gen_var_prin}.  Using the definition of the interactions
(\ref{eq:genKcov}), the covariance of a system of $N+M$ spins
$\gamma=(\alpha,\sigma)$   is given by
\begin{equation}
\Ev{ K_{N+M}( \gamma) K_{N+M}( \gamma')} \, = \, \frac{N+M}{2} \,
f(q_{\gamma, \gamma'}),
\end{equation}
where we have adopted the notation used in
Section~\ref{cav_perspect}. To first order, the overlap of the
combined system may be expressed in terms of the overlaps within
the two blocks as
\begin{equation}
q_{\gamma, \gamma'} \, = \, q_{\alpha, \alpha'} \, + \, \left( \,
q_{\sigma,
    \sigma'} \, - \, q_{\alpha, \alpha'} \, \right) \,
    \frac{M}{N} + O\Big(\Big(\frac{M}{N}\Big)^2\Big)\, ,
\end{equation}
see equation (\ref{eq:decqtt}). Taylor expansion of the function
$f$, again to first order, yields
\begin{equation} \label{eq:gendec}
\frac{N+M}{2} \, f( q_{\gamma, \gamma'}) \, - \, \frac{N}{2} \, f(
q_{\alpha, \alpha'}) \, = \, - \, \frac{M}{2} \phi( q_{\alpha,
  \alpha'}) \, + \, \frac{M}{2} \, q_{\sigma, \sigma'} \, f'(
q_{\alpha, \alpha'})+ O\Big( \frac{M^2}{N} \Big)\, ,
\end{equation}
where
\begin{equation} \label{eq:defphi}
\phi(q) \, := \, q \, f'(q) \, - \, f(q).
\end{equation}
Now, given a ROSt $\mu$, one may define two sets of independent,
centered gaussian random variables $\{ \kappa_{\alpha} \}$ and $\{
V_{\alpha, \sigma} \}$, which are attuned to the more general
Hamiltonian (\ref{eq:genham}). As indicated by (\ref{eq:gendec}),
these random variables are defined by prescribing their
covariances as follows:
\begin{equation} \label{eq:genkapcov}
\E \left( \kappa_{\alpha} \, \kappa_{\alpha'} \right) \, = \,
\frac{\phi( q_{\alpha, \alpha'})}{2}\, ,
\end{equation}
where $\phi$ is as defined in (\ref{eq:defphi}), and
\begin{equation} \label{eq:genVcov}
\E \left( V_{\alpha, \sigma} \, V_{\alpha', \sigma'} \right) \, =
\, \frac{M}{2} \, f'( q_{\alpha, \alpha'}) \, q_{\sigma,
\sigma'}\, ,
\end{equation}
(the positivity of the covariance can be concluded  from the
representation \eqref{genint}, \cite{AS2}). Correspondingly, a
free energy functional, analogous to (\ref{eq:Gdef}), may be
defined as
\begin{equation} \label{eq:genfef}
G_M( \beta, h; \mu) = \frac{1}{M} \E \left( \ln \left[ \frac{
\sum_{
      \alpha, \sigma} \xi_{\alpha} e^{ \beta \left( V_{\alpha, \sigma}
      + \underline{h} \cdot \underline{ \sigma} \, \right) }}
      { \sum_{\alpha} \xi_{\alpha} e^{ \beta \sqrt{M}
    \kappa_{\alpha}}} \right] \right).
\end{equation}

With these new definitions, one may derive a variational principle
analogous to Theorem~\ref{thm:as2}.  Moreover, as in
Theorem~\ref{thm:as2}, ROSts formed by  $N$ particle systems with
the  Hamiltonian \eqref{eq:genham} may be used to demonstrate that
the inequality actually saturates.  Through an adaptation of the
methods discussed above one can prove:

\begin{thm} \label{thm:genas2} Let $\beta \geq 0$ and $h \in \R$.
\newline i) For any $M \in \mathbb{N}$,
\begin{equation} \label{eq:genub}
P_M( \beta, h) \leq \inf_{\mu:{\rm ROSt}} G_M( \beta, h; \mu)\, .
\end{equation}
ii) The pressure of the system corresponding to (\ref{eq:genham})
may be realized through:
\begin{equation} \label{eq:genvarprin}
P( \beta,h) = \lim_{M \to \infty} \inf_{\mu:{\rm ROSt}} G_M(
\beta, h; \mu).
\end{equation}
\end{thm}

For further discussion the reader is referred to \cite{AS2}.   

\section*{Acknowledgement}
We thank J. C. Zambrini for his encouragement and the invitation
to write this summary.   Some of the work was done at the Weizmann
Institute, where MA has enjoyed the hospitality of the Department
of Physics of Complex Systems.  The work was supported in part by
NSF Grant DMS-0602360  and, at its early stages, by NSF
Postdoctoral Fellowships (RS, SLS).

 \end{document}